\documentclass[apj]{emulateapj}
\usepackage[colorlinks,linkcolor={blue},citecolor={blue},urlcolor={red}]{hyperref}

\let\counterwithin\relax

\usepackage{epsfig,graphicx,natbib,amsmath,amsfonts,amssymb}
\usepackage{color}
\usepackage[dvipsnames]{xcolor}
\usepackage{multirow}
\usepackage{booktabs}
\usepackage{amssymb}
\usepackage{threeparttablex} 
\usepackage{booktabs}
\usepackage{chngcntr}
\usepackage{longtable}
\usepackage{array} 

\newcommand{\fq}{$f_{\rm Q}$}

\newcommand{\Msunh}{\>h^{-1}\rm M_\odot}
\newcommand{\msolar}{M$_\odot$}
\newcommand{\mstar}{$M_\ast$}
\newcommand{\lgmstar}{$\log_{10}$($M_\ast/h^{-2}$\msolar)}

\newcommand{\logmhalo}{\log(M_{\rm h}/h^{-1}\Msun)}

\newcommand{\Msun}{{\rm  M}_{\odot}}

\newcommand{\myemail}{\email{ecwang16@ustc.edu.cn(EW), whywang@ustc.edu.cn(HW)}}

\shorttitle{The dearth of difference between central and satellite galaxies}
\shortauthors{Wang et al.}

\graphicspath{{fig/}}

\begin{document}

\title{The dearth of difference between central and satellite galaxies II. Comparison of observations with L-GALAXIES and EAGLE in star formation quenching}

\author{
Enci Wang\altaffilmark{1,2},
Huiyuan Wang\altaffilmark{1,2},
Houjun Mo\altaffilmark{3,4},
Frank C. van den Bosch\altaffilmark{5},
S.H. Lim\altaffilmark{4},
Lixin Wang\altaffilmark{3},
Xiaohu Yang\altaffilmark{6,7},
Sihan Chen\altaffilmark{1}
} \myemail

\altaffiltext{1}{CAS Key Laboratory for Research in Galaxies and Cosmology, Department of Astronomy, University of Science and Technology of China, Hefei 230026, China}
\altaffiltext{2}{School of Astronomy and Space Science, University of Science and Technology of China, Hefei 230026, China}
\altaffiltext{3}{Tsinghua Center of Astrophysics \& Department of Physics, Tsinghua University, Beijing 100084, China}
\altaffiltext{4}{Department of Astronomy, University of Massachusetts, Amherst MA 01003-9305, USA}

\altaffiltext{5}{Department of Astronomy, Yale University, P.O. Box 208101, New Haven, CT 06520-8101, USA}
\altaffiltext{6}{Department of Astronomy, Shanghai Jiao Tong University, Shanghai 200240, China}
\altaffiltext{7}{IFSA Collaborative Innovation Center, Shanghai Jiao Tong University, Shanghai 200240, China}


\begin{abstract}
  As we demonstrated in Paper I, the quenched fractions of central and
  satellite galaxies as function of halo mass are extremely similar,
  as long as one controls for stellar mass. The same holds for the
  quenched fractions as a function of central velocity dispersion,
  which is tightly correlated with black hole mass, as long as one
  controls for both stellar and halo mass.  Here we use mock galaxy
  catalogs constructed from the latest semi-analytic model,
  L-GALAXIES, and the state-of-the-art hydrodynamical simulation,
  EAGLE, to investigate whether these models can reproduce the trends
  seen in the data.  We also check how the group finder used to
  identify centrals and satellites impacts our results. We find that
  L-GALAXIES fails to reproduce the trends.  The predicted quenched
  fraction of central galaxies increases sharply with halo mass around
  $10^{12.5}h^{-1}\Msun$ and with black hole mass around
  $\sim10^{6.5}$\msolar, while the predicted quenched fraction of
  satellites increases with both halo and black hole masses
  gradually. In contrast, centrals and satellites in EAGLE follow
  almost the same trend as seen in the data. We discuss the
  implications of our results for how feedback processes regulate
  galaxy quenching.

\end{abstract}


\section{Introduction}
\label{sec:introduction}

Star formation quenching is an essential process in galaxy formation
and evolution since redshift about 2.5 \citep[e.g.][]{Bundy-06,
  Faber-07, Muzzin-12, Tomczak-14, Barro-17}. However, what causes the
star formation to quench is still an unsolved problem. Many
observational studies have found that the quenched fraction of
galaxies depends strongly on other intrinsic properties of galaxies,
such as stellar mass, bulge-to-total light-ratio and central stellar
velocity dispersion \citep[e.g.][]{Driver-06, Cameron-09, Wuyts-11,
  Mendel-13, Fang-13, Bluck-14, Woo-15, Wang-18b}, as well as on the
environments they reside in \citep[e.g.][]{Balogh-04, vandenBosch-07,
  Peng-10, Peng-12, Woo-13, Wang-16, Wang-18}.

Central and satellite galaxies in dark matter halos, as represented
observationally by galaxy groups \citep[e.g.][]{Yang-05}, are usually
assumed to experience different quenching processes, because of their
different locations and motions within the host halos.  This is
supported by observations that the quenched fraction of centrals is
found to strongly depend on their structures, but only weakly on their
host halo mass \citep{Fang-13, Bluck-14, Woo-15} and environmental
density \citep[e.g.][]{Peng-10, Peng-12}, while the quenched fraction
of satellites appears to be more sensitive to their environments
\citep{Abadi-Moore-Bower-99, Balogh-Navarro-Morris-00,
  Blanton-Roweis-07, vandenBosch-08, Weinmann-09, Peng-10, Wolf-09,
  Woo-13}. In fact, the differences in the quenching properties
between centrals and satellites are usually used to estimate the
efficiency of satellite-specific quenching processes
\citep[e.g.][]{vandenBosch-08, Wetzel-Tinker-Conroy-12, Knobel-13,
  Kovac-14, Knobel-15, Wang-18}.


However, there is growing evidence that centrals and satellites are
not as different as expected. For example, when dividing galaxies into
several narrow stellar mass bins, \cite{Hirschmann-14} found that the
quenched fractions of centrals and satellites respond to their local
densities in a similar way.  \cite{Knobel-15} found that, if
satellites have similar stellar mass as the centrals, the two
populations have on average the same quenched fraction. Similarly,
\cite{Bluck-16} found that the quenching properties of massive
satellites are identical to those of centrals of the same stellar
mass.  \cite{Wang-18} analyzed the quenched fractions in samples
controlled by both stellar mass and halo mass, and found that
satellites and centrals in those controlled samples have similar
quenching properties.  They demonstrated that the different quenching
behaviours of centrals and satellites found previously are mainly
due to the fact that satellites, on average, reside in more massive
halos than centrals of the same stellar mass. In particular, although
centrals and satellites of similar stellar masses (or residing in
halos of similar masses) reside in halos of different halos mass (or
have different stellar mass distributions), their quenched fractions
follow the same correlation with stellar mass (or halo mass).

More recently, in the first paper of this series, \citet[][hereafter
  Paper I]{WangE-18} presented a comprehensive analysis on the
quenching properties of centrals and satellites based on the SDSS
group catalog of \citep{Yang-07, Yang-Mo-vandenBosch-09}, including the
quenched fraction as a function of stellar mass, bulge-to-total light
ratio, central velocity dispersion, halo mass and halo-centric radius.
The results show that the central and satellite populations exhibit
similar trends of quenched fraction with all the variables
considered.  Moreover, we found that the fractions of Seyfert and
radio galaxies, which relate to different types of AGN feedback, are
also the same for centrals and satellites, as long as the stellar and
halo mass are controlled. This suggests that the quenching mechanisms
working on centrals and satellites are related to these internal and
environmental parameters in a similar way.

Understanding how, when, and where galaxies quench their star
formation is one of the most muddled, outstanding problems in galaxy
formation.  Different models for galaxy formation and evolution often
make very different assumptions and/or predictions. Among them,
semi-analytic galaxy formation models \citep[SAM;
  e.g.][]{White-Frenk-91, Kang-05, Bower-06, Croton-06,
  Bower-McCarthy-Benson-08, Somerville-08, Parry-Eke-Frenk-09, Lu-11,
  Guo-11, Lee-14, Lu-14a, Gonzalez-Perez-14,
  Somerville-Popping-Trager-15, Ruiz-15, Henriques-15} and
hydro-dynamical simulations \citep[e.g.][]{Katz-Hernquist-Weinberg-92,
  Springel-01, Springel-Hernquist-03, Springel-05, Keres-09,
  Angulo-12, Vogelsberger-14a, Vogelsberger-14b, Schaye-15} are two
powerful tools to trace galaxy formation and evolution in cosmological
volumes.  SAMs are phenomenological models that use
approximate/empirical formula to describe all baryonic processes
relevant to galaxy formation, such as gas accretion, heating and
cooling, star formation, feedback from stars and AGN, mergers, and
stripping due to tides and ram pressure.  Because of their different
locations within host halos, centrals and satellites in SAMs are
assumed to undergo different quenching processes \citep[see
  e.g.][]{Henriques-17}. For example, the stripping of hot and cold
gas associated with galaxies is assumed to only act on satellites.  In
addition, the efficiency of radio AGN feedback is assumed to depend on
the associated hot gas mass, which may be very different between
centrals and satellites. Hydro-dynamical simulations, on the other
hand, evolve the dark matter and baryonic components in a
self-consistent way, though some assumptions have also to be adopted
to model subgrid physics \citep[see e.g.][]{Springel-Hernquist-03}.
More importantly, centrals and satellites are not treated differently
{\it a priori}, and their differences, if any, result directly from a
complicated interaction between the baryonic content of a galaxy and
its environment, while the (subgrid) modelling of star formation and
feedback processes carry no knowledge of this environment.

In general, the current models are able to reproduce many global
properties of observed galaxies, such as their abundance, their
clustering, and the overall trend of the quenched fraction as a
function of galaxy environment and galaxy properties (see the model
papers cited above). However, they usually have difficulties in
matching details of the quenched fraction \citep{Hirschmann-14,
  Somerville-Dave-15, Khandai-15, Furlong-15, Guo-16,
  Henriques-17}. This indicates that improvements in the treatments of
quenching processes in semi-analytic models and in the sub-grid
prescriptions in hydro-dynamic simulations are still required.


In this paper, we compare the observational results obtained in Paper
I with theoretical predictions, based on both the L-GALAXIES
semi-analytical galaxy formation model of \citet{Henriques-15}, and
the state-of-the-art hydro-dynamic simulation EAGLE \citep[Evolution
  and Assembly of GaLaxies and their Environments;][]{Schaye-15}.  In
particular, we focus on whether the models can reproduce the
similarity among centrals and satellites seen in the data.  The
comparisons between observations and models and between different
models on the central-satellite difference may provide valuable
insight into quenching mechanisms, thereby helping to improve galaxy
formation models. Since the halo masses and central/satellite
classification in the data are based on the group finder of
\citet{Yang-05}, which carries its own uncertainties, it is
important to take this into account when comparing models and/or
simulations to data. In particular, \cite{Campbell-15} have shown that
the red fraction of central galaxies as function of halo mass is not
well recovered by the group finder.  We therefore construct mock data
sets from L-GALAXIES and EAGLE, and run the same group finder over
these mock data sets in order to facilitate a more meaningful
comparison.

The remainder of this paper is organized as follows. In Section
\ref{sec:mock} we present the galaxy formation models and the mock
catalogs constructed from them. In Section \ref{sec:models} we
compare the observational results in Paper I with predictions from
L-GALAXIES and EAGLE.  In Section \ref{sec:gf} we examine the
uncertainties that may be caused by the group finder. We discuss the
implications of our results in Section \ref{sec:summary}.

\section{Galaxy Formation Models and Mock catalogs}\label{sec:mock}


Semi-analytic galaxy formation models and hydro-dynamic simulations,
which take into account various baryonic processes, are the most
powerful tools to study galaxy formation and evolution.  Here we use
data from the latest version of the Munich model,
L-GALAXIES\footnote{http://galformod.mpa-garching.mpg.de/public/LGalaxies}
\citep{Henriques-15, Henriques-17} and the state-of-the-art
hydrodynamic simulation,
EAGLE\footnote{http://eagle.strw.leidenuniv.nl/}
\citep[][]{Schaye-15,Crain-15,Furlong-15}, to compare with the
observational results we have obtained in Paper I.

\subsection{L-GALAXIES and EAGLE}

The L-GALAXIES model used here, as described in detail in
\citet{Henriques-15}, is an updated version of the Munich
semi-analytic model \citep[e.g.][]{Croton-06, Guo-11} built upon the
Millennium Simulation \citep{Springel-05} assuming the Planck
cosmology: $\Omega_{m}=0.315$, $\Omega_{\Lambda}=0.685$ and $h=0.673$
\citep{Planck-14b}. L-GALAXIES employs a Markov Chain Monte Carlo
(MCMC) method to explore the high-dimensional parameter space to match
the observed galaxy abundance and quenched fraction as a function of
stellar mass from redshift of 3 down to 0. To better match the
observational results, \cite{Henriques-15} made some changes in the
model, including delaying the reincorporation of galaxy wind ejecta,
eliminating the ram-pressure stripping in small halos, and modifying
the model for radio mode AGN feedback.  Due to the mass resolution
limits of the Millennium Simulation, L-GALAXIES does not include
galaxies with \mstar$<$10$^{9.5}$\msolar.

In L-GALAXIES, two major processes are assumed to play important roles
in quenching star formation (or maintaining a low star formation rate)
in passive galaxies: environmental effects on satellites, and radio
mode feedback from central supermassive black holes. Heating from
radio mode feedback is set to be proportional to the black hole mass
and the hot gas mass in the halo (for centrals) or subhalo (for
satellites) in question.  Thus, the quenched fraction predicted by
L-GALAXIES is directly linked to the mass of the hot gas and that of
the central black hole, in the sense that galaxies with more hot gas
and a more massive central black hole are more likely to be quenched.
For satellites, additional ``satellite specific" quenching processes
are taken into account. After a satellite falls into a massive galaxy
cluster, stripping of gas due to tidal interactions and ram pressure
is assumed to remove its hot and cold gas reservoir gradually,
subsequently leading to star formation quenching \citep[see
  also][]{Gunn-Gott-72, Toomre-Toomre-72, Moore-96,
  Boselli-Gavazzi-06, Read-06, Balogh-Navarro-Morris-00, Weinmann-09,
  Wang-15}.  In the current version of L-GALAXIES
\citep{Henriques-15}, the ram-pressure is assumed to remove the hot
gas of satellites only in massive halos with $M_{\rm
  h}>10^{14}$\msolar. In addition, the gas surface density threshold
for turning cold gas into stars is reduced by a factor of almost two
in comparison with the value assumed earlier. Here we use the snapshot
at $z=0.1$, which roughly corresponds to the median redshift of our
SDSS galaxy sample (see Paper I). The simulation box is 480.3 Mpc/h on
a side.

EAGLE consists of a number of hydrodynamic simulations that follow the
formation and evolution of galaxies and supermassive black holes in
the $\Lambda$ cold dark matter universe. The adopted cosmology is also
the Planck cosmology, but with slightly different parameters from that
of L-GALAXIES: $\Omega_{m}=0.307$, $\Omega_{\Lambda}=0.693$ and
$h=0.678$ \citep{Planck-14a}.  These simulations adopt advanced
smoothed particle hydrodynamics and subgrid physical models for gas
cooling, metal enrichment, black hole growth, and stellar and AGN
feedback. Free parameters in the feedback models are calibrated using
the galaxy stellar mass function and the stellar mass - black hole
mass relation at $z\sim0$ \citep{Crain-15, Furlong-15}. The
\cite{Chabrier-03} initial mass function and the
\cite{Bruzual-Charlot-03} stellar population model are used to obtain
luminosities and stellar masses of individual galaxies based on their
star formation histories. In a simulation like EAGLE, the same subgrid
physical prescriptions are used for both centrals and satellites, and
the differences between galaxies come directly from the formation and
evolution processes.  For example, the AGN feedback is assumed to
depend on the local gas properties rather than on the gas properties
of the whole halo.  Similarly, the stripping processes of hot and cold
gas associated with galaxies are directly modeled from first
principles, independent of whether a galaxy is a central or satellite.
In this paper, we use the simulation Ref-L100N1504, which has a box
size of 100 Mpc and uses 2$\times1504^3$ particles. The masses of gas
and dark matter particles are 1.81$\times10^6$\msolar\ and
9.70$\times10^6$\msolar, respectively. The simulation contains more
than 11,000 dark matter halos with masses above 10$^{11}$\msolar, and
nearly 10,000 galaxies with masses comparable to or larger than that
of the Milky Way.

The cosmological parameters used in L-GALAXIES and EAGLE are not
exactly the same as the ones used in Paper I. However, the differences
are sufficiently small that this should not affect any of our results
regarding the quenching properties of centrals and satellites. Another
subtle difference regards the definition of centrals and
satellites. In EAGLE, centrals are defined as the most massive
galaxies in their host halos, with all other galaxies being satellites
\citep{Furlong-15}. In L-GALAXIES, on the other hand, a central galaxy
is defined as the galaxy that resides at the center of the main
progenitor halo, which is the progenitor that evolves along the main
trunk of the halo merger tree. This is not necessarily the most
massive galaxy within that halo. However, we find that more than 97\%
of all centrals identified in L-GALAXIES are the most massive
galaxies. Therefore, this subtle difference in central/satellite
classification will not significantly impact any of our results.

\subsection{Mock catalogs and definition of the quenched population}
\label{sec:mockcat}

Our goal is to investigate the quenching properties of centrals and
satellites for L-GALAXIES and EAGLE, and examine whether or not the
models can reproduce the trends seen in the SDSS data.  To make
reliable comparisons with the observations, we construct mock catalogs
of galaxies for both L-GALAXIES and EAGLE, so as to take into account
observational selection effects \citep[for details see][]{Lim-17}. In
both case, we first stack the duplicates of the original simulation
box side by side to construct a sufficiently large volume. We then
choose a location for the observer in the constructed volume, and
calculate the redshift and the apparent magnitude for each galaxy
based on its luminosity, distance and movement with respect to the
observer.  Finally, we select a flux-limited sample of galaxies from a
light cone covering the redshift range $0.01<z<0.2$, which is similar
to the range covered by the SDSS data.  All comparisons are based on
the mock catalogs, unless specified otherwise. Since the volume of the
EAGLE simulation is quite small ($100\times100\times100 {\rm Mpc}^3$),
there are many repeated sources in its mock catalog. Therefore, the
error bars of the statistics for massive galaxies and halos may be
significantly underestimated in the EAGLE sample.

In Section \ref{sec:models}, we use the mock galaxy catalogs thus
constructed, together with the halo masses and central/satellite
classification taken directly from the models, to analyze the
correlations of the quenching properties with various other galaxy and
environmental parameters.  We refer to these mock catalogs as
L-GALAXIES and EAGLE, respectively. In Section \ref{sec:gf}, we will
take into account the uncertainties induced by the group finder.  We
apply the same group finder as used for the SDSS data to the two mock
galaxy catalogs to select galaxy groups and to assign halo masses to
them according to their characteristic, total stellar mass, following
the method developed in \cite{Yang-05, Yang-07}. For clarity, we refer
to the mock galaxy catalogs with halo masses and central/satellite
classification as inferred from the group finder as L-GALAXIES+GF and
EAGLE+GF, respectively.

For our analysis, we use a number of quantities provided by L-GALAXIES
and EAGLE, including star formation rate (SFR), stellar mass ($M_*$),
host halo mass ($M_{\rm h}$), the mass of the central supermassive
black hole ($M_{\rm BH}$), and the scaled halo-centric radius ($R_{\rm
  p}/r_{\rm 180}$), defined as the projected distance from the galaxy
to the host group center in units of the halo virial radius of the
host group. To compare with the results in Paper I, we treat the
luminosity-weighted average of the positions of member galaxies in a
group as the group center, and define the group virial radius as the
radius within which the dark matter halo has an overdensity of 180
\citep[see Equation 5 in][]{Yang-07}.

In paper I, we adopted the demarcation line suggested by
\cite{Bluck-16} to separate galaxies into star-forming and quenched
populations.  In Appendix A, we show the specific SFR-\mstar\ diagram
for SDSS, L-GALAXIES and EAGLE. As one can see, the results for the
two models are quite different from the SDSS
data. 
Thus, adopting the same demarcation line for the two models as for the SDSS
galaxies is inappropriate.  Instead, for L-GALAXIES/EAGLE, we first divide the
galaxies into 15 stellar mass bins from $10^9h^{-2}$\msolar\ to
$10^{12}h^{-2}$\msolar. We then sort the galaxies in each bin by their
SFRs, and set the SFR threshold below which the model galaxies are
considered to be quenched such that the resulting quenched fraction as
a function of stellar mass for the two models are exactly the same as
that for the SDSS galaxies. The SFR threshold as a function of stellar mass
for both L-GALAXIES and EAGLE are presented in Appendix A. 

The quenched fraction ($f_Q$) for the mock catalogs is calculated in
the following way by using the $V_{\rm max}$ method to correct the
Malmquist bias.  For a given subsample (S), the quenched fraction
(\fq) is defined as:
\begin{equation}\label{eq:qf}
 f_{\rm Q}=\frac{\sum_{i=1}^{N_S} w_i\times f_{\rm Q,i}}{\sum_{i=1}^{N_S} w_i}.
\end{equation}
Here $N_S$ is the number of galaxies in subsample S, $w_{i}=1/V_{\rm
  max}$ is the weight given to galaxy $i$, and $f_{\rm Q,i}$
represents the corresponding quenched status: $f_{\rm Q,i}=1$ if the
galaxy is quenched, otherwise $f_{\rm Q,i}=0$.  As in Paper I, the
error in the quenched fraction is estimated by using 1000 bootstrap
samples.

\section{Comparisons of L-GALAXIES and EAGLE with SDSS data}
\label{sec:models}

In this section, we investigate whether the similarity between
centrals and satellites seen in Paper I for SDSS galaxies can be
reproduced by L-GALAXIES and EAGLE.

We emphasize that the quenched populations for the two models are redefined so that a meaningful comparison between models and observational data can be made. In Appendix A, we also use an identical threshold to separate galaxies into star-forming and quenched populations for both the observation and the models. As one can see, both models can reproduce the overall trend of the quenched fraction increasing with stellar mass. However they both underestimate the quenched population, and the results for EAGLE are even worse than those for L-GALAXIES. These results are broadly consistent with those obtained before by, e.g., \citet{Furlong-15} and 
\citet{Henriques-17}. We refer the reader to these papers for details.
In the following, we adopt the new definition for quenched population (see Section \ref{sec:mockcat} and Appendix A) and analyze the
quenched fraction as a function of galaxy stellar mass, halo mass and
central black hole mass separately for centrals and satellites in the
L-GALAXIES and EAGLE catalogs (see Section \ref{sec:mock}), and
compare the results with the SDSS results obtained in Paper I. Note
that the halo masses and central/satellite classification used here
are the true values taken directly from the models.

\subsection{Dependence of the quenched fraction on stellar mass and halo mass}


\begin{figure*}
 \begin{center}
  \epsfig{figure=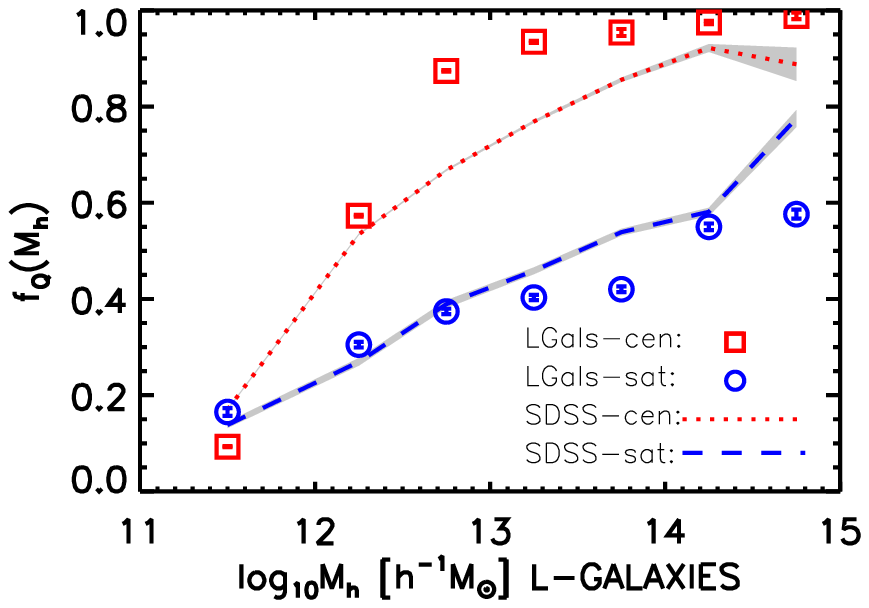,clip=true,width=0.4\textwidth}
  \epsfig{figure=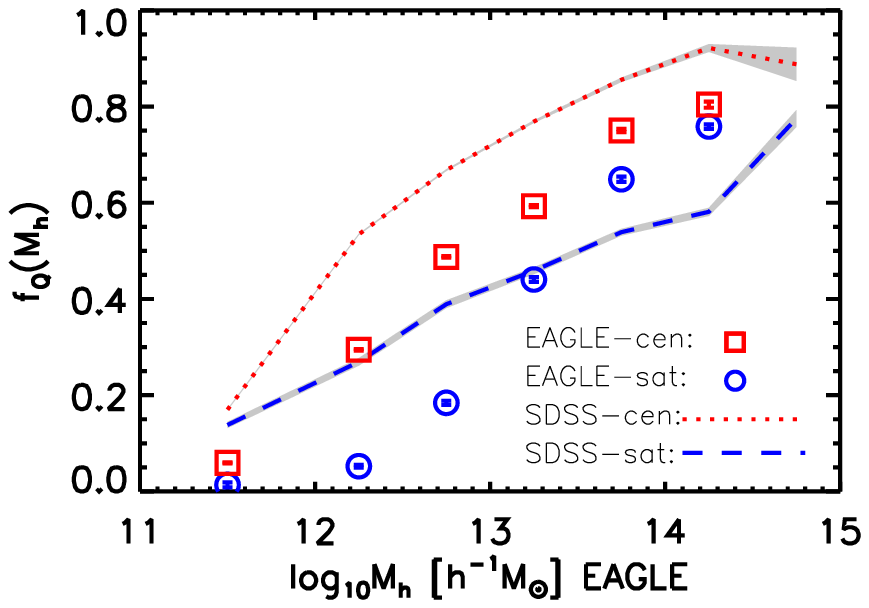,clip=true,width=0.4\textwidth}
 \end{center}
  \caption{Quenched fraction of mock galaxies as a function of halo
    mass for L-GALAXIES (left panel) and EAGLE (right panel)
    catalogs. The red squares and blue circles indicate the results
    for centrals and satellites, respectively. For comparison, we also
    show the results for SDSS galaxies (dotted and dashed lines with
    error bars represented by the shaded region, taken from the left
    panel of figure 2 in Paper I). }
 \label{fig:SAM_quench_mh_0th}
\end{figure*}

\begin{figure*}
  \begin{center}
    \epsfig{figure=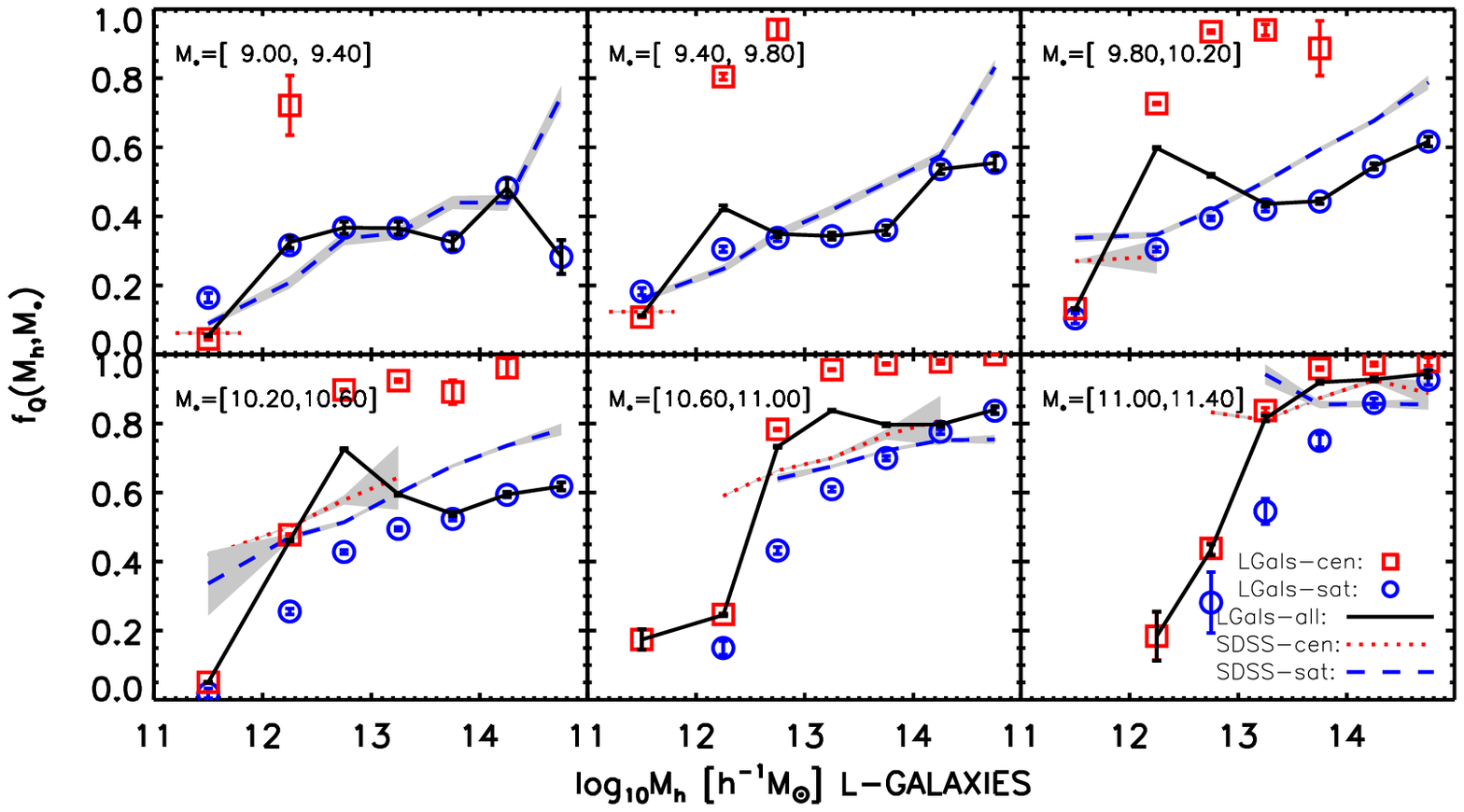,clip=true,width=0.7\textwidth}
    \epsfig{figure=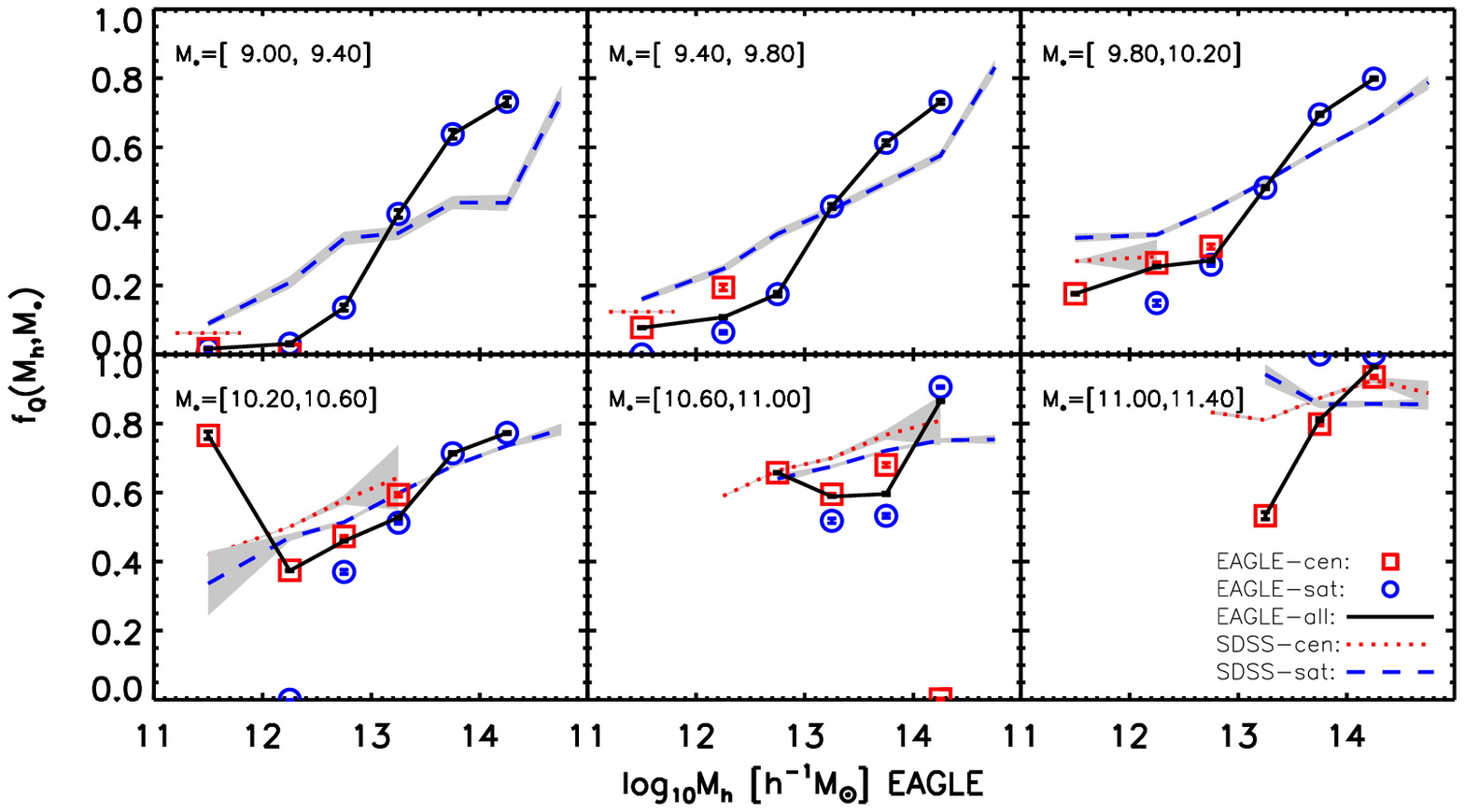,clip=true,width=0.7\textwidth}
  \end{center}
\caption{Quenched fraction of mock galaxies as a function of halo mass
  for central (red squares), satellite (blue circles), and all
  galaxies (black lines) in different stellar mass bins, as indicated
  in each panel.  The upper and lower groups of panels show the
  results for L-GALAXIES and EAGLE, respectively. The results for SDSS
  centrals and satellites, taken from the top panels of figure 3 in
  Paper I, are shown by the red dotted and blue dashed lines,
  respectively.  }
 \label{fig:SAM_quench_mh_1th}
\end{figure*}

In Figure \ref{fig:SAM_quench_mh_0th}, we present the \fq-$M_{\rm h}$
relations for centrals (red squares) and satellites (blue circles)
based on L-GALAXIES (left panel) and EAGLE (right panel). For
comparison, the results for SDSS galaxies, taken from the left panel
of figure 2 in Paper I, are also shown.  As one can see, both
L-GALAXIES and EAGLE are able to reproduce the overall trends of
\fq\ with $M_{\rm h}$ for both the centrals and satellites.  At fixed
halo mass, the centrals are more frequently quenched than satellites,
consistent with the observational data. For L-GALAXIES, the
\fq-$M_{\rm h}$ relations are in fairly good agreement with the data,
except at intermediate halo mass ($\sim 10^{13}h^{-1}$\msolar), where
the predicted quenched fraction of centrals is significantly higher
than observed. In contrast, centrals in EAGLE are less frequently
quenched than in the SDSS almost over the entire halo mass range.
Moreover, satellites in EAGLE appear to be more sensitive to halo mass
than what is seen in the data. In general, as far as the overall
\fq-$M_{\rm h}$ relations are concerned, L-GALAXIES is in better
agreement with the SDSS data than EAGLE.

To see the dependence of \fq\ on $M_{\rm h}$ in more detail, we divide
the galaxies into six stellar mass bins, and show the \fq-$M_{\rm h}$
relations for individual stellar mass bins in Figure
\ref{fig:SAM_quench_mh_1th}.  For comparison, the corresponding
results for SDSS galaxies, taken from the top panels of figure 3 in
Paper I, are also presented.  Compared with SDSS, the satellites in
L-GALAXIES exhibit weaker dependence of the quenched fraction on halo
mass at low stellar masses, but stronger dependence at high stellar
masses. For centrals, on the other hand, the predicted dependence by
L-GALAXIES is stronger than that in the SDSS over the entire stellar mass
range. More importantly, at given stellar mass the \fq-$M_{\rm h}$
relations for centrals and satellites are different, contrary to what
is seen in the SDSS data. The quenched fraction of central galaxies
increases rapidly with halo mass around $M_{\rm
  h}\sim10^{12.5}h^{-1}$\msolar\ and is close to unity for
$M_{\rm h} > 10^{13} h^{-1}$\msolar. This trend appears without strong
dependence on stellar mass.  In contrast, the quenched fraction of
satellites increases gradually with halo mass and the halo mass
dependence strengthens with increasing stellar mass. This is in good
agreement with the results of \cite{Hirschmann-14}, who used the SAM
of \citet{Guo-11} and found that centrals are more frequently quenched
than satellites at fixed local density when galaxies are divided into
a series of narrow stellar mass bins.  Since the two populations
exhibit very different dependence on halo mass and reside in different
halos at given stellar mass, a cusp appears in the \fq-$M_{\rm h}$
relation for all galaxies, in disagreement with the observational
results \citep[e.g.][Paper I]{Peng-12, Hirschmann-14, Woo-15,
  Wang-18}.

For EAGLE, the difference of centrals and satellites shown in Figure
\ref{fig:SAM_quench_mh_0th} is reduced when stellar mass is
controlled, as shown in Figure \ref{fig:SAM_quench_mh_1th}. EAGLE
galaxies show a steeper dependence on $M_{\rm h}$ than SDSS galaxies
at \mstar$<10^{10.2}h^{-2}$\msolar, but overall the agreement with the
data is fairly good. Note that there are only 233 and 30 galaxies in
the original simulation box for the two largest stellar mass bins,
respectively. Hence, these results carry a large uncertainty, which is
not properly captured by the errorbars shown, due to the fact that the
mock catalog is constructed using repeated stacking of the EAGLE
simulation volume.

These results clearly show that L-GALAXIES fails to reproduce the
similarity in quenching properties between centrals and satellites.
EAGLE, on the other hand, yields \fq-$M_{\rm h}$ relations that reveal
significant discrepancies with respect to the data, but it does
predict that centrals and satellites have similar quenching properties
when controlled for both stellar and halo mass.

\subsection{The \fq-M$_{\rm BH}$ relation at fixed halo and stellar masses}
\label{subsec:fqbh}

\begin{figure*}
  \begin{center}
    \epsfig{figure=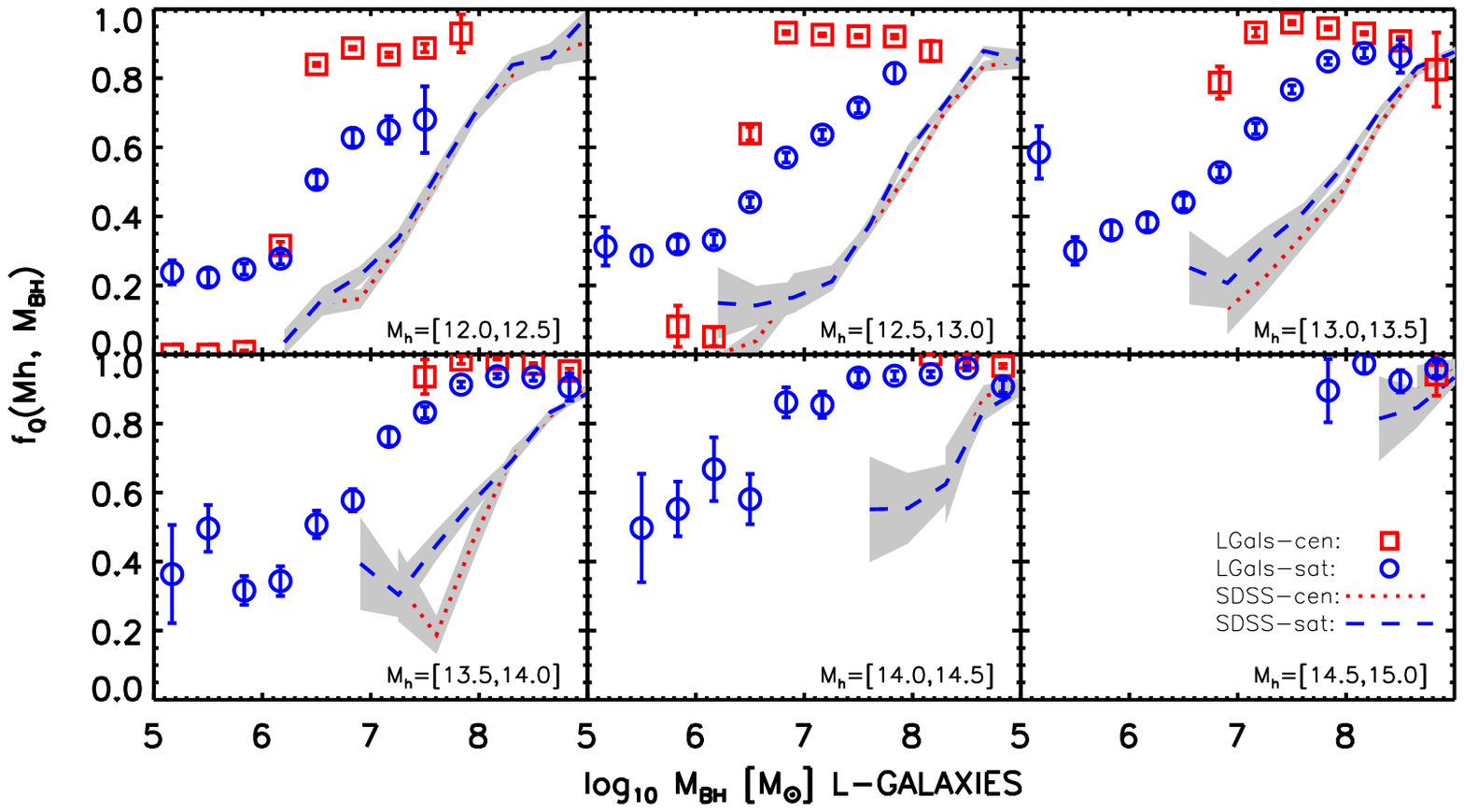,clip=true,width=0.7\textwidth}
    \epsfig{figure=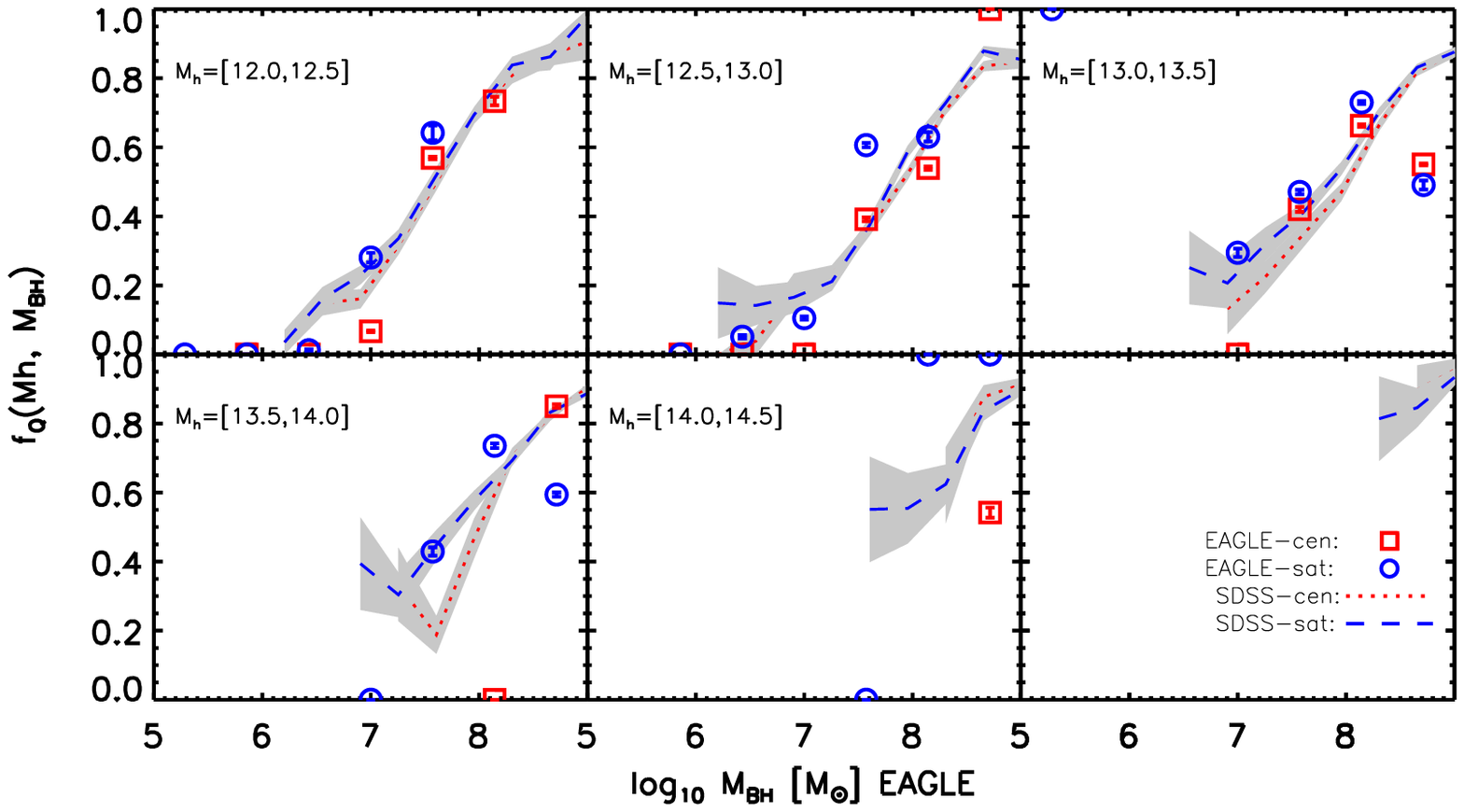,clip=true,width=0.7\textwidth}
  \end{center}
\caption{Top group of panels: quenched fraction as a function of central black hole
mass for centrals (red solid lines) and satellites (blue solid lines) predicted by
L-GALAXIES in different halo mass and stellar mass ranges. Bottom group of panels:
the same as the top group of panels, but for the predictions by
EAGLE. The lack of halos with $M_{\rm h}>10^{14.5}h^{-1}$\msolar\ is due to the small
volume of the EAGLE simulation. For comparison, we also present the results for SDSS
galaxies (red dotted line for centrals and blue dashed line for satellites) in each
panel. The SDSS results are derived from Paper I by adopting the
$M_{\rm BH}$-$\sigma_c$ relation from \cite{Saglia-16}. The halo mass range
is labeled in each panel, and the stellar mass bins can be found in
Appendix B.}
\label{fig:SAM_quench_bh_2th}
\end{figure*}


Several other galaxy properties, such as the bulge-to-total light
ratio, bulge mass, and central velocity dispersion, are found to be
better correlated with the quenched fraction than stellar mass
\citep[e.g.][]{Driver-06, Bell-08, Wuyts-11, Mendel-13, Fang-13,
  Woo-15, Wang-18b}.  More recently, using observations of 91 galaxies
with directly measured black hole masses, \cite{Terrazas-16} found
that quenched galaxies have more massive black holes than star-forming
galaxies of similar stellar mass. Consistent with this, the central
velocity dispersion of a galaxy, which is tightly correlated with
black hole mass, also correlates with whether a galaxy is quenched or
not \citep{Bluck-16, Teimoorinia-Bluck-Ellison-16}. It is thus
interesting to compare the predicted \fq-$M_{\rm BH}$ relations of
centrals and satellites in L-GALAXIES and EAGLE, both among each other
and against observational data.  Since the black hole masses for SDSS
galaxies are not available, we follow \cite{Bluck-16} and estimate the
black hole masses by using the well established $M_{\rm
  BH}$-$\sigma_{\rm c}$ relation \citep[e.g.][]{Ho-08, Kormendy-Ho-13,
  McConnell-Ma-13}.  Specifically, we use the relation, $\log_{\rm
  10}(M_{\rm BH}[M_{\odot}]) = 5.25\times \log_{\rm 10}(\sigma_{\rm c}
[\rm km\ s^{-1}]) - 3.77$, obtained by \citet{Saglia-16}, which
has an observed scatter of about 0.46 dex.

Figure \ref{fig:SAM_quench_bh_2th} shows the \fq-$M_{\rm BH}$
relations of centrals (red dashed line) and satellites (blue dashed
line) for L-GALAXIES (top group of panels) and EAGLE (bottom group of
panels) in a series of halo mass bins. For comparison, the
corresponding results for SDSS galaxies are also presented (the same
results as shown in figure 7 of Paper I, but with $\sigma_{\rm c}$
replaced with $M_{\rm BH}$).  In order to eliminate the effects of
stellar mass, the centrals and satellites shown in each panel are
restricted to a narrow stellar mass bin. For each halo mass bin, the
stellar mass bin is selected to make sure that we have sufficiently
many centrals and satellites (see Appendix B).  Thus, the stellar mass
bin chosen varies with halo mass. Moreover, the stellar mass
distributions for SDSS, L-GALAXIES and EAGLE galaxies are different,
and so the stellar mass bins are slightly different for them, even for
a given halo mass bin (see Appendix B). Since the difference is small
and our purpose is to compare centrals and satellites, this does not
affect any of our conclusions.

As shown in Figure \ref{fig:SAM_quench_bh_2th}, L-GALAXIES reproduces
the overall trend in the data, in that the quenched fraction increases
with $M_{\rm BH}$. However, there are discrepancies. At 
low $M_{\rm BH}$, the quenched fractions at given black hole mass 
predicted by the model are much higher than in the data. 
Our tests show that these discrepancies are caused by the fact 
that the black hole masses predicted by L-GALAXIES are systematically 
lower than those for SDSS galaxies of the same stellar mass \citep[see
  also][]{Terrazas-16}. In addition, the predicted \fq-$M_{\rm BH}$
relation for centrals show a clear jump, with the quenched fraction
increasing rapidly from close to zero at $M_{\rm BH}< 10^{6.5}\Msun$
to almost unity above $10^{7}\Msun$. The results for satellites are very 
different, particularly for galaxies in low mass halos. The \fq-$M_{\rm BH}$ 
relation of satellites is smooth and does not show
the jump around $M_{\rm BH}\sim10^{6.5}\Msun$ seen for centrals.
We emphasize, however, that the systematic underprediction of black hole mass
is unlikely the main cause for the difference between centrals and satellites, 
as the problem with black hole mass appears in both populations. 
We suspect that the difference may be related to the different efficiency in 
AGN fueling and feedback, as discussed in Section \ref{sec:summary}.

In contrast to L-GALAXIES, the range of the black hole masses
predicted by EAGLE is similar to that of SDSS galaxies at given halo
and stellar mass.  This is largely expected, as the feedback
efficiency in EAGLE has been calibrated to match the amplitude of the
observed galaxy-black hole mass relation. In the three low halo mass
bins, the predicted \fq-$M_{\rm BH}$ relations are in good agreement
with the SDSS data.  Most importantly, the relations for centrals and
satellites are very similar, consistent with what is seen in the
data. For more massive halos, the results show large fluctuations, due
to the small numbers of galaxies in the corresponding mass bins.

In summary, L-GALAXIES and EAGLE predict very different \fq-$M_{\rm
  BH}$ relations for both centrals and satellites. Whereas the EAGLE
results are in excellent agreement with the SDSS data, the trends
predicted by L-GALAXIES are extremely discrepant.  We will discuss the
implications of these results in Section \ref{sec:summary}.

\subsection{Dependence of the \fq-\mstar\ relation on halo mass and halo-centric distance}

\begin{figure*}
  \begin{center}
    \epsfig{figure=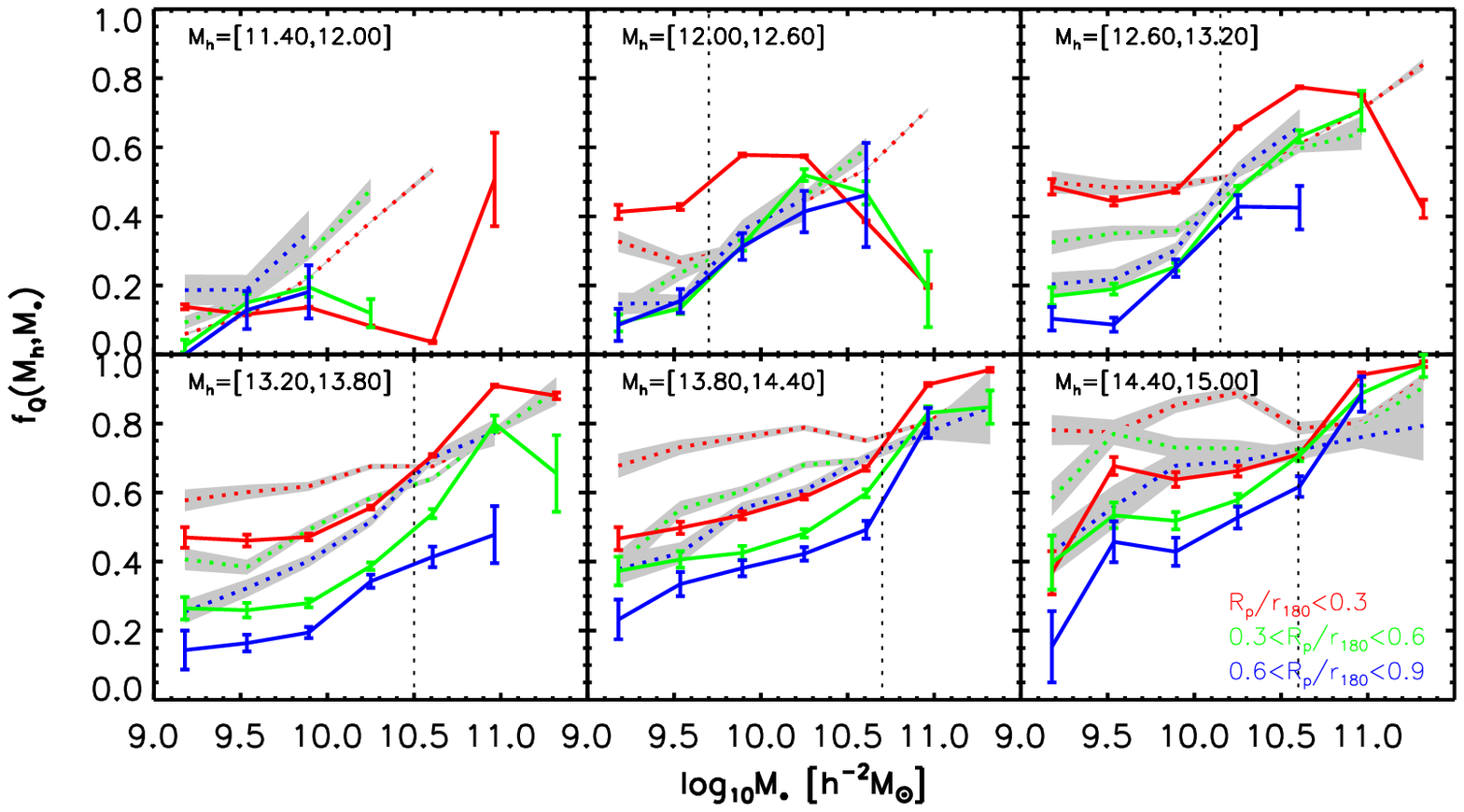,clip=true,width=0.7\textwidth}
    \epsfig{figure=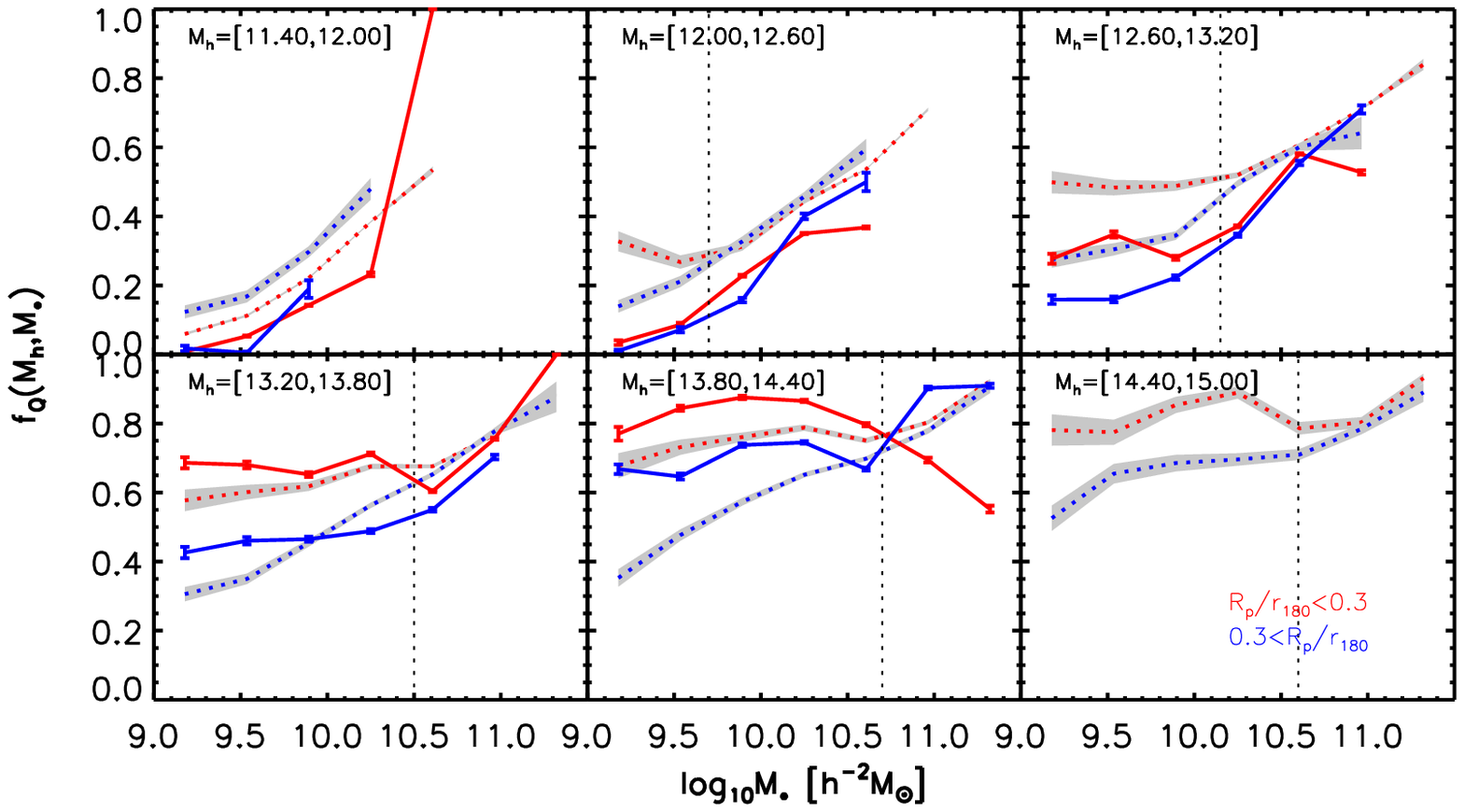,clip=true,width=0.7\textwidth}
  \end{center}
\caption{Top group of panels: quenched fraction as a function of
  stellar mass in the six halo mass bins as predicted by
  L-GALAXIES. Results are shown for galaxies in three halo-centric
  distance intervals: $R_{\rm p}$/$r_{180}<$0.3 (red solid lines),
  $0.3<R_{\rm p}$/$r_{180}<0.6$ (green solid lines), and $0.6<R_{\rm
    p}$/$r_{180}<0.9$ (blue solid lines).  Bottom group of panels: the
  same as the top panels but for the predictions of EAGLE. Here
  results are shown for two halo-centric distance intervals: $R_{\rm
    p}$/$r_{\rm 180}<0.3$ (red solid lines) and $0.3<R_{\rm
    p}$/$r_{\rm 180}$ (blue solid lines). For comparison, for each
  halo mass bin, we plot the quenched fraction as a function of
  stellar mass for SDSS galaxies divided by the halo-centric distance,
  which are the same as what we have done for L-GALAXIES and EAGLE.
These are taken from figure 9 of Paper I. The shaded regions are the 1$\sigma$
confidence range. The dotted vertical lines indicate the stellar mass threshold,
above which the \fq\ of galaxies appears to be independent of halo-centric
distance at given halo mass. }
 \label{fig:fq_mstar_radius}
\end{figure*}

Satellite galaxies are usually assumed to experience a number of
`satellite-specific' quenching processes, such as tidal interaction
\citep[e.g.][]{Toomre-Toomre-72, Read-06}, ram-pressure stripping
\citep[e.g.][]{Gunn-Gott-72, Abadi-Moore-Bower-99, Hester-06, Wang-15}
and strangulation \citep{Larson-80, Balogh-Navarro-Morris-00,
  vandenBosch-08}.  All these may lead to dependence of \fq\ on
halo-centric distance.  Indeed, satellites of a given stellar mass are
found to be more frequently quenched near group/cluster centers than
in the outer parts \citep[e.g.][]{Weinmann-06, vandenBosch-08b,
  Wetzel-Tinker-Conroy-12, Kauffmann-13, Wang-18}.  In Paper I, we
found that the value of \fq\ depends on halo-centric distance
significantly only for galaxies with low masses, and that there seems
to be a stellar mass threshold for given halo mass, above which the
quenched fraction becomes independent of halo-centric distance.  In
this subsection, we test whether or not L-GALAXIES and EAGLE can
reproduce this observational result.

Figure \ref{fig:fq_mstar_radius} shows the \fq-\mstar\ relation in the
six halo mass bins, as predicted by L-GALAXIES (top group of panels)
and by EAGLE (bottom group of panels), with each sample further
divided according to halo-centric distance. For comparison, we also
present similar results for SDSS galaxies (taken from Paper I) in
dotted lines with shaded regions. The dotted vertical lines indicate
the stellar mass threshold derived from the observational results,
above which the quenched fraction becomes independent of halo-centric
distance. We divide L-GALAXIES galaxies into three halo-centric
distance bins, and EAGLE galaxies into two bins to achieve better
statistics. Since centrals and satellites of similar stellar mass show
very similar dependence on halo-centric distance, at least in the data
(see Paper~I), we do not investigate centrals and satellites
separately.

Both L-GALAXIES and EAGLE can successfully reproduce the trend that
galaxies located closer to the group center are more frequently
quenched over almost the entire halo mass range.  However,
above the stellar mass thresholds, indicated by the vertical lines, the 
quenched fraction predicted by L-GALAXIES still shows significant dependence 
on halo-centric distance, at least for intermediate halo mass bins
($10^{12.0}h^{-1}$\msolar$<M_{\rm h}<10^{13.8}h^{-1}$\msolar),
which is inconsistent with the observational result.
In contrast, EAGLE appears to match the data
much better, at least in the intermediate halo mass range from
$10^{12.6}h^{-1}$\msolar\ to $10^{13.8}h^{-1}$\msolar.  Note that no
EAGLE results are plotted for the halo mass bin
$10^{14.4}h^{-1}M_{\odot} < M_{\rm h} < 10^{15.0}h^{-1}M_{\odot}$, due the
the small number of massive galaxies in the EAGLE simulation
(see Figure~\ref{fig:halo_mass}).

\section{Uncertainties induced by the group finder algorithm}\label{sec:gf}

\begin{figure*}
  \begin{center}
    \epsfig{figure=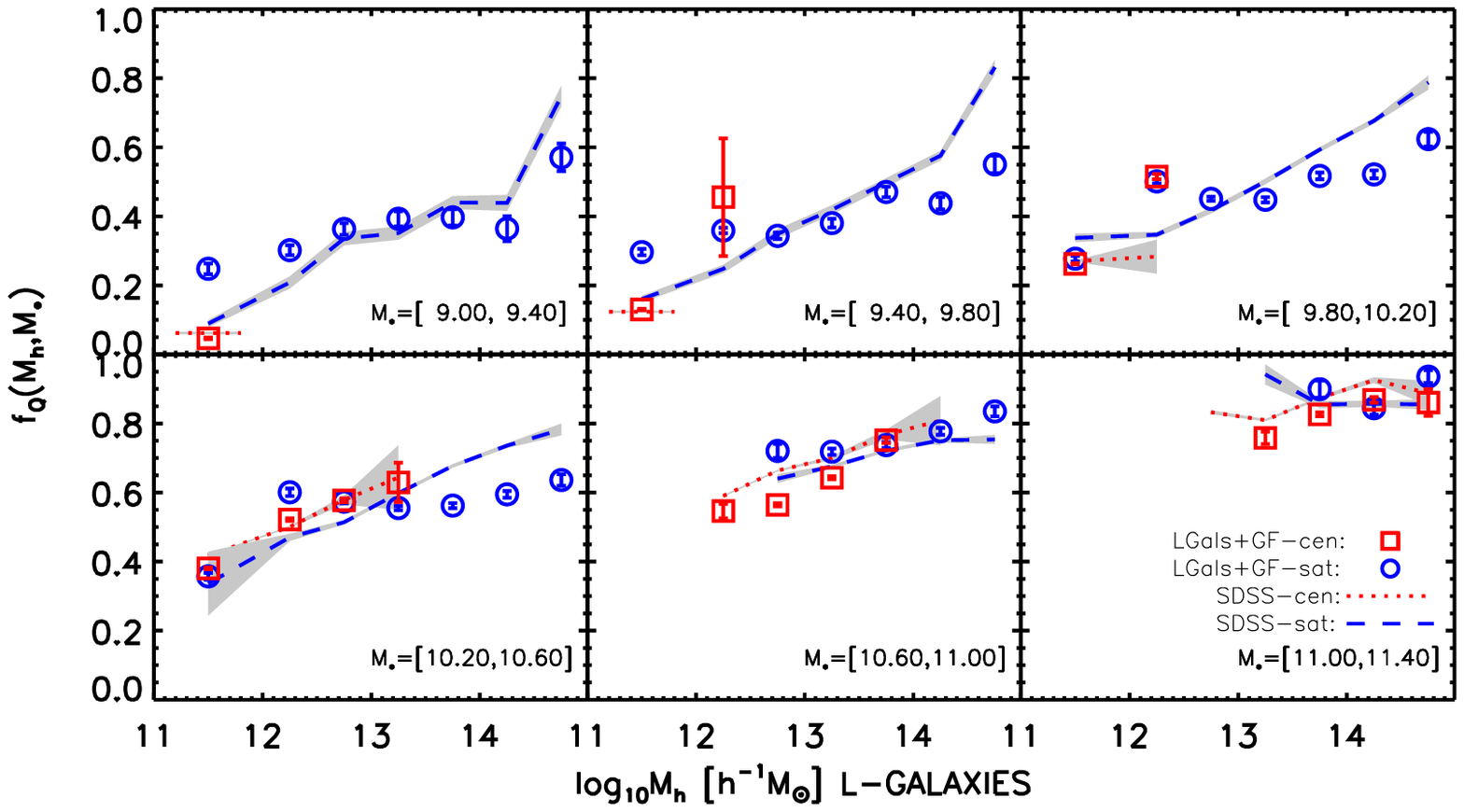,clip=true,width=0.7\textwidth}
    \epsfig{figure=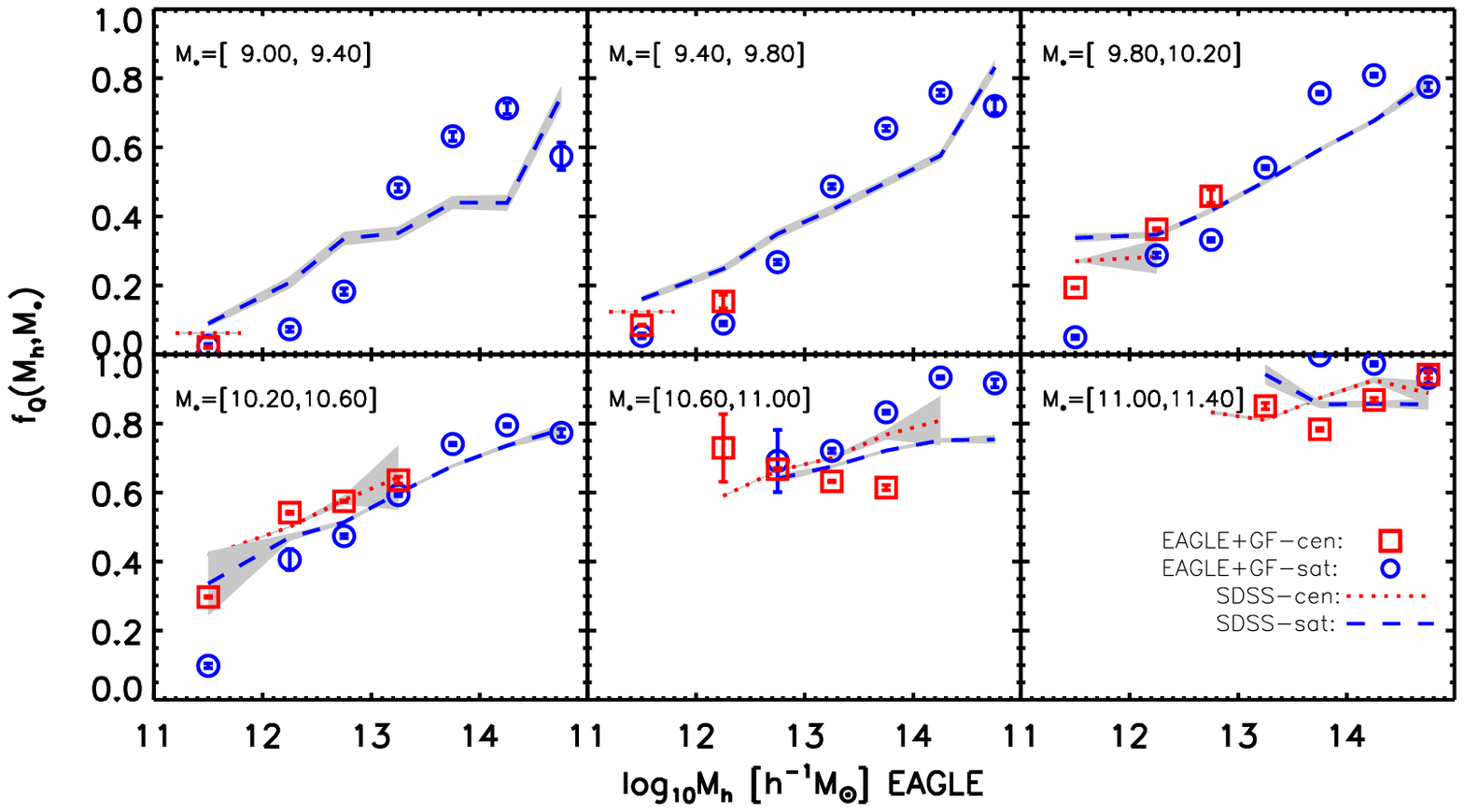,clip=true,width=0.7\textwidth}
  \end{center}
\caption{The same as Figure \ref{fig:SAM_quench_mh_1th}, but obtained from
the mock samples, L-GALAXIES+GF  (top group of panels) and EAGLE+GF
(bottom group of panels).
 }
 \label{fig:SAM_quench_mh_1th_GF}
\end{figure*}

\begin{figure*}
  \begin{center}
    \epsfig{figure=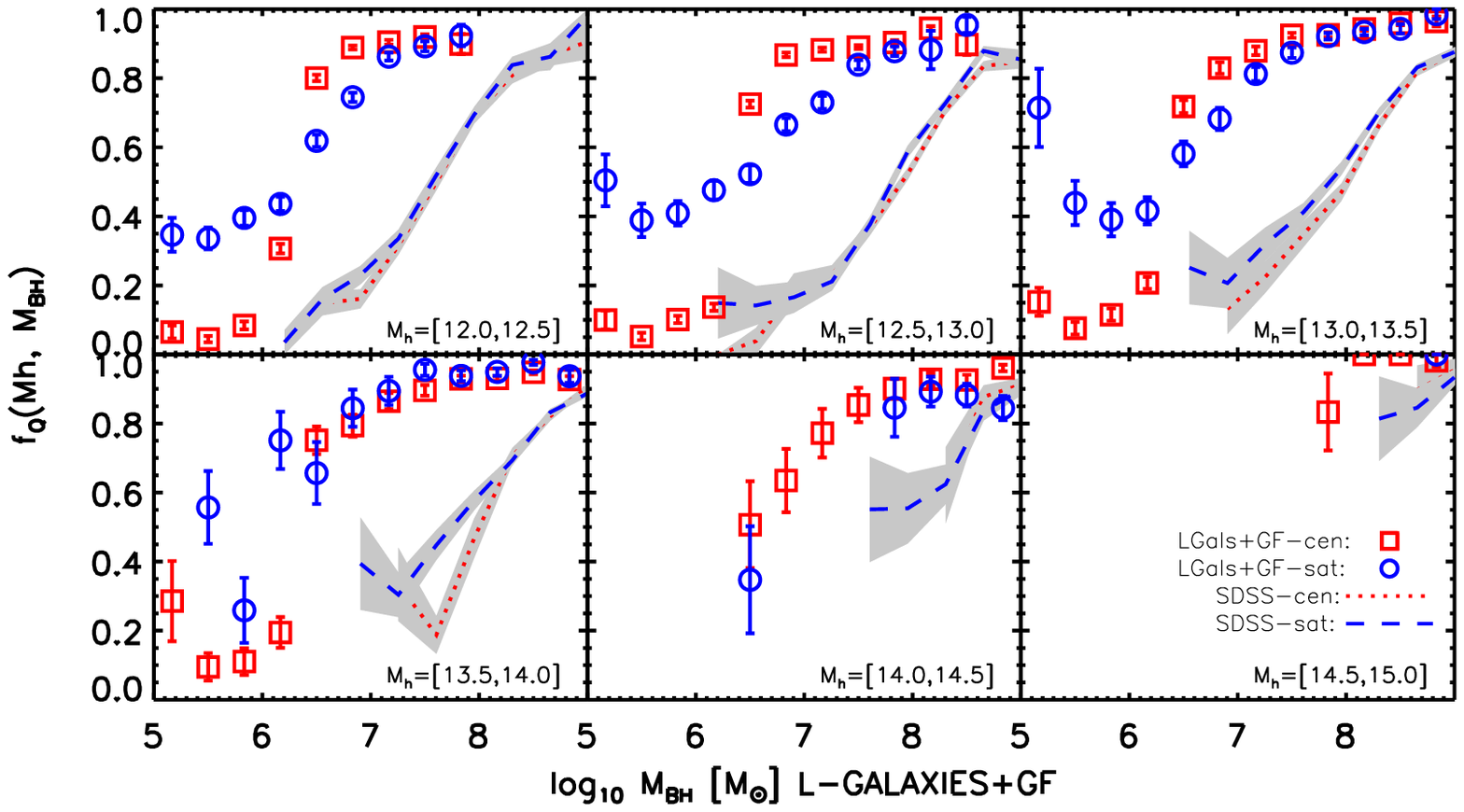,clip=true,width=0.7\textwidth}
    \epsfig{figure=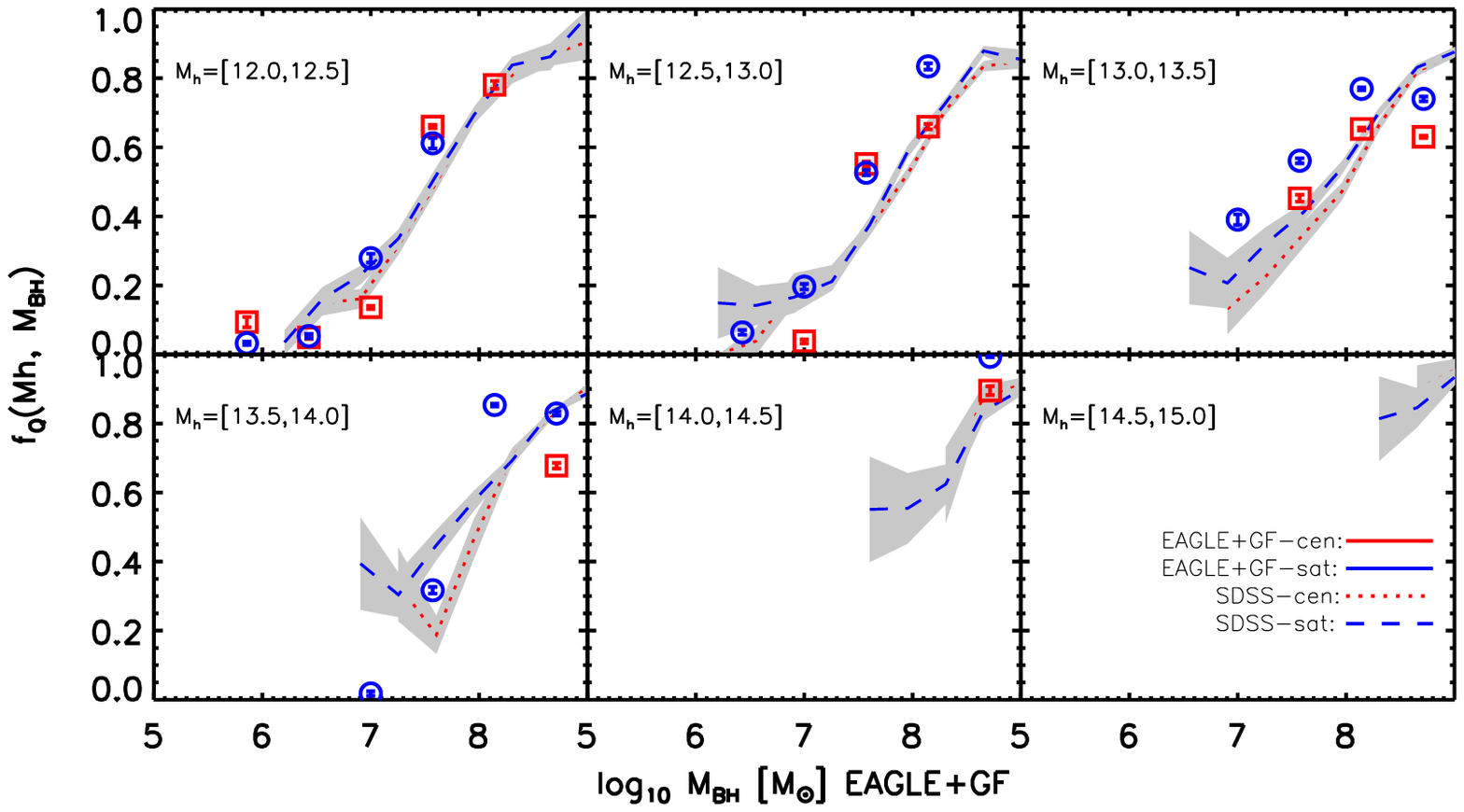,clip=true,width=0.7\textwidth}
  \end{center}
\caption{The same as Figure \ref{fig:SAM_quench_bh_2th}, but obtained from
the mock samples, L-GALAXIES+GF  (top group of panels) and EAGLE+GF (bottom group of panels).
}
\label{fig:SAM_quench_bh_2th_GF}
\end{figure*}

The SDSS results shown above and in Paper I are based on the galaxy
group catalog constructed using the halo-based group finder developed
by \citet{Yang-05, Yang-07}. The results presented in the previous
section do not take into account uncertainties that may be induced by
the group finder. In particular, the group finder assigns halo masses
to each group based on the group's total, characteristic stellar mass
\citep[see][for details]{Yang-07}, and characterizes galaxies as
centrals or satellites based on their stellar mass rank within the
group, with centrals being the brightest group members. This
invariably comes with some errors. In a recent study,
\cite{Campbell-15} has shown that these errors typically reduce the
differences between centrals and satellites, making them appear more
similar than they really are. In this section, we investigate the
reliability of our results against potential uncertainties introduced
by the group finder.

\subsection{Dependence on stellar mass, halo mass, black hole mass
and halo-centric distance}\label{subsec:gf_fq}

To do this, we use the L-GALAXIES+GF and EAGLE+GF catalogs, described
in Section \ref{sec:mock}, to derive the statistics of model galaxies.
Note that, for these two catalogues, the halo masses and
central/satellite classification are obtained from the group finder.
The upper and lower sets of panels in Figure
\ref{fig:SAM_quench_mh_1th_GF} show the quenched fractions as a
function of halo mass for galaxies in the same six stellar mass bins
as before, as obtained from L-GALAXIES+GF and EAGLE+GF,
respectively. The SDSS results are also plotted for comparison.

For L-GALAXIES, the group finder clearly has a dramatic impact
  on the quenched fractions. In particular, the large differences
  between centrals and satellites evident in Figure~
  \ref{fig:SAM_quench_mh_1th} are now significantly reduced.  However,
  the \fq-$M_{\rm h}$ relations for centrals and satellites obtained
  from L-GALAXIES+GF are still significantly different for the two
  lowest stellar mass bins, $[9,9.4]$ and $[9.4,9.8]$. Moreover, the
  group finder tends to make the \fq-$M_{\rm h}$ relations flatter for
  both centrals and satellites. In particular, for satellites in halos
  with $M_{\rm h}>10^{12}h^{-1}$\msolar, the quenched fraction is
  almost independent of halo mass. In contrast, the quenched fraction
  of the observed satellites reveals a strong mass dependence.

The group finder has a much weaker impact on the results of the EAGLE
simulation.  As shown in the lower set of panels in
Figure \ref{fig:SAM_quench_mh_1th_GF}, the \fq-$M_{\rm h}$ relations
obtained from EAGLE+GF for centrals and satellites have only changed
slightly compared to the results shown in
  Figure \ref{fig:SAM_quench_mh_1th}. In particular, the impact of the
group finder on the quenched fraction is similar for centrals and
satellites, and the flattening effect that plagues the L-GALAXIES
results is not significant for EAGLE.

The fact that the group finder significantly reduces the
  difference among the \fq-$M_{\rm h}$ relations of centrals and
  satellites for the L-GALAXIES model is somewhat unfortunate. It
  makes the quenching properties of L-GALAXIES and EAGLE appear more
  similar than they really are. As a result, the shortcomings of
the group finder prevent us from clearly preferring one model over the
other, at least when it comes to the difference in quenching
statistics for centrals and satellites that are controlled for both
halo mass and stellar mass. However, as we will see below, when it
comes to the black hole mass dependence and the dependence on halo-centric
radius, EAGLE+GF clearly outperforms L-GALAXIES+GF.

The \fq-$M_{\rm BH}$ relations obtained from L-GALAXIES+GF and
EAGLE+GF are shown in Figure \ref{fig:SAM_quench_bh_2th_GF} for six
halo mass bins.  As in Section \ref{subsec:fqbh}, centrals and
satellites are restricted to narrow stellar mass bins (see Appendix B)
to eliminate the dependence on stellar mass. For the L-GALAXIES model,
applying the group finder does not change the results too much; the
large difference between centrals and satellites can still be seen for
$M_{\rm h}<10^{13.5}h^{-1}M_{\odot}$.  In particular, the strong jump
of \fq\ for centrals at $M_{\rm BH}\sim10^{6.5}\Msun$ does not change
significantly from L-GALAXIES to L-GALAXIES+GF. For the EAGLE model,
the group finder algorithm also has only weak impact on the
\fq-$M_{\rm BH}$ relations of centrals and satellites, although the
uncertainties are large, particularly for more massive halos.

The reason why the \fq-$M_{\rm BH}$ relation is less sensitive
  to the group finder than the \fq-$M_{\rm h}$ relation, is simply
  because $M_{\rm BH}$ does not enter in, or derive from, the group
  finder, whereas $M_{\rm h}$ does; the only way that the group finder
  can impact the \fq-$M_{\rm BH}$ relations is through
  mis-classification of centrals and satellites. The results presented
  here suggest that this is not a major source of error. The
  implication is, that the similarity in the relations between \fq\ and
  central velocity dispersion for centrals and satellites presented in
  Paper~I is not an artifact of the group finder. Since central
  velocity dispersion is strongly correlated with black hole mass, we
  argue that the SDSS data is in much better agreement with EAGLE than
  with L-GALAXIES.


\begin{figure*}
  \begin{center}
    \epsfig{figure=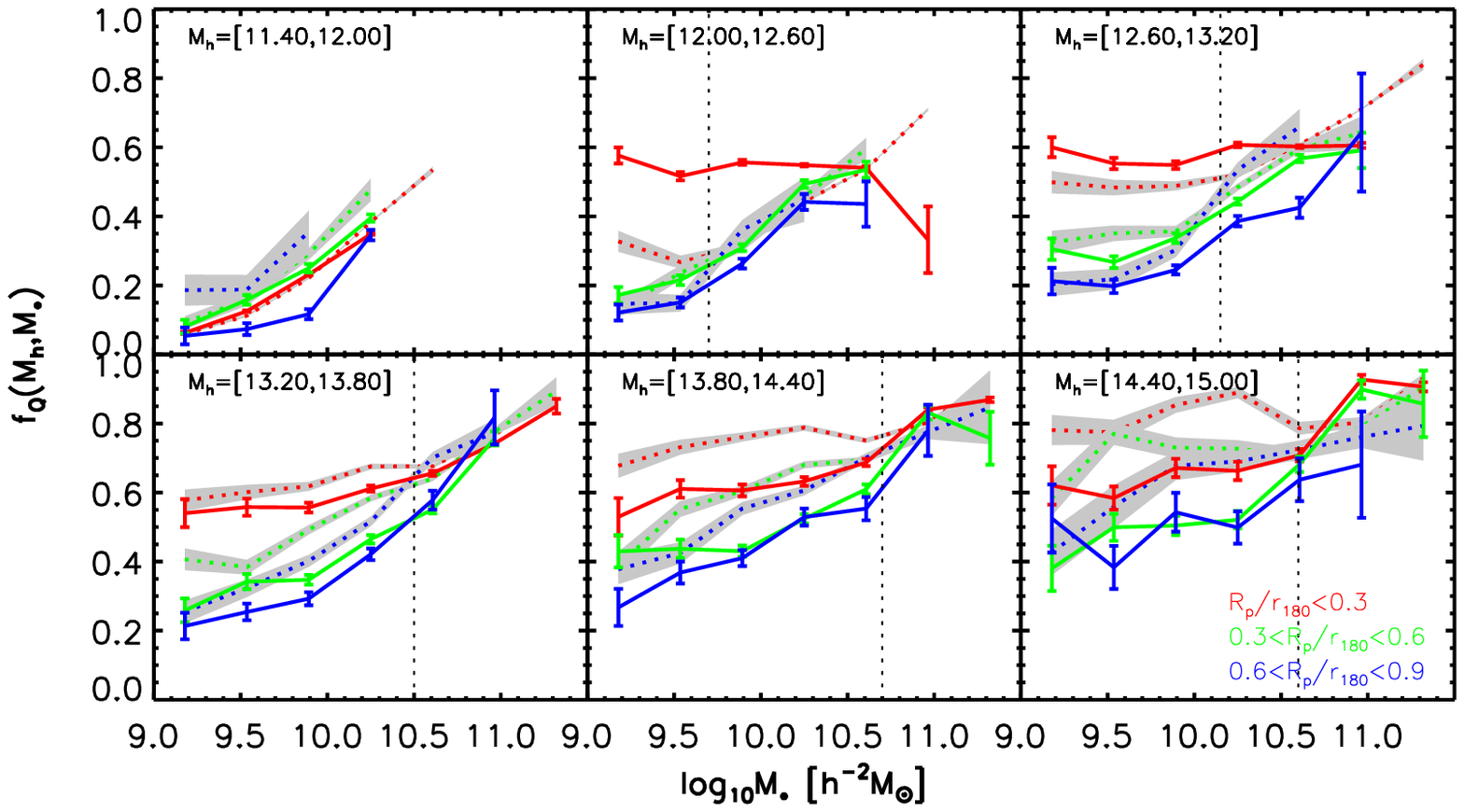,clip=true,width=0.7\textwidth}
    \epsfig{figure=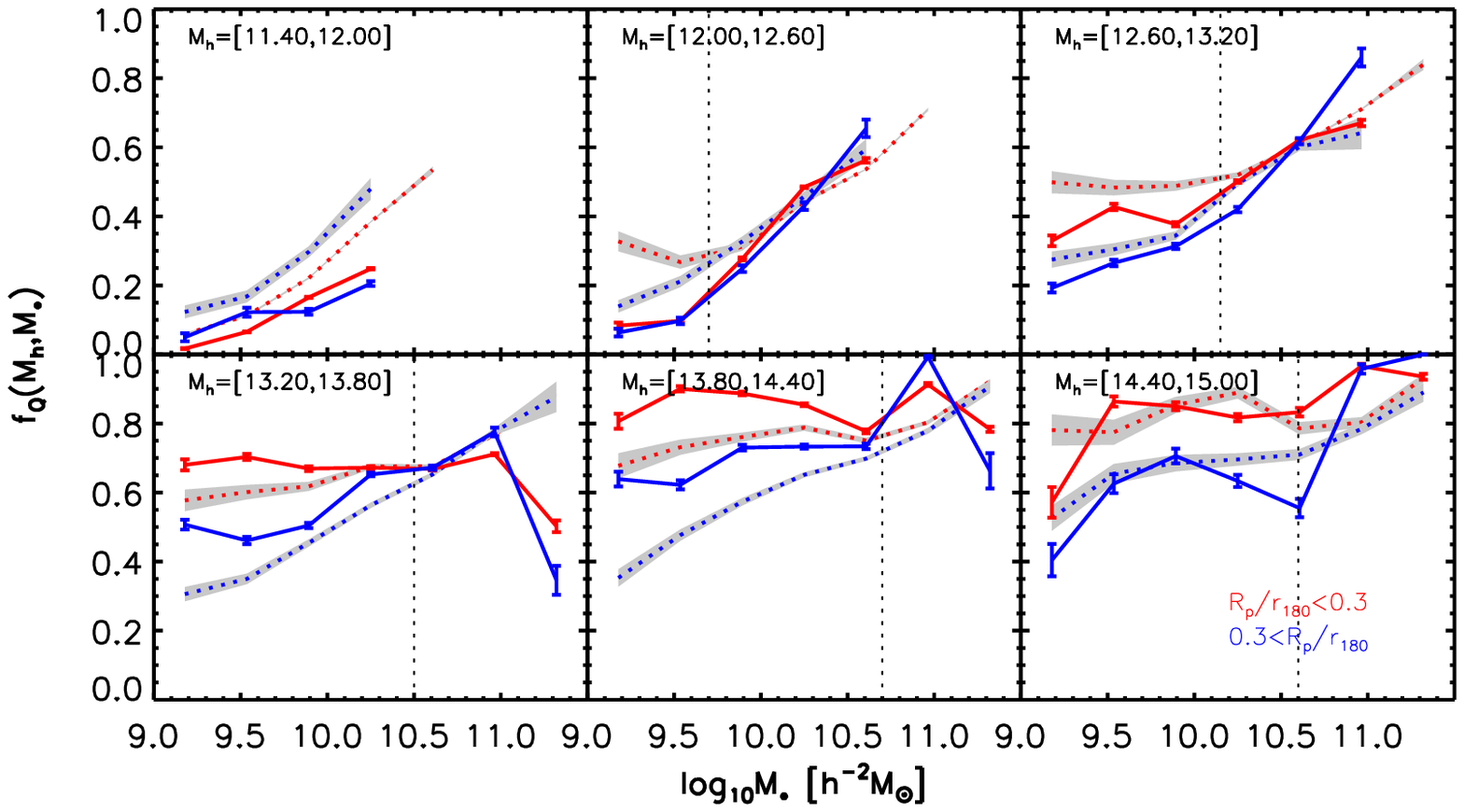,clip=true,width=0.7\textwidth}
  \end{center}
\caption{The same as Figure \ref{fig:fq_mstar_radius},  but obtained from the mock samples, L-GALAXIES+GF (top group of panels) and EAGLE+GF (bottom group of panels).}
 \label{fig:fq_mstar_radius_gf}
\end{figure*}

Finally, Figure \ref{fig:fq_mstar_radius_gf} presents the
\fq-\mstar\ relations in different halo-centric distance intervals
predicted by L-GALAXIES+GF and EAGLE+GF.  For L-GALAXIES, the
application of the group finder changes some details in the
relationship, but the overall trends remain.  For massive halos,
  with $M_{\rm h} > 10^{13.2} \Msunh$, the results from L-GALAXIES+GF
  are now in better agreement with the SDSS results. However, for
  halos with $10^{12.0} \Msunh < M_{\rm h} < 10^{13.2} \Msunh$, the
  significant dependence of \fq\ on halo-centric radius is still
  apparent for galaxies above the stellar mass threshold indicated by
  the dotted, vertical lines.  For EAGLE, the group finder has little
  impact, and the \fq-\mstar\ relations are very similar to the
  results shown in Figure \ref{fig:fq_mstar_radius}.


\subsection{Uncertainties introduced by the group finder}

\begin{figure*}
  \begin{center}
    \epsfig{figure=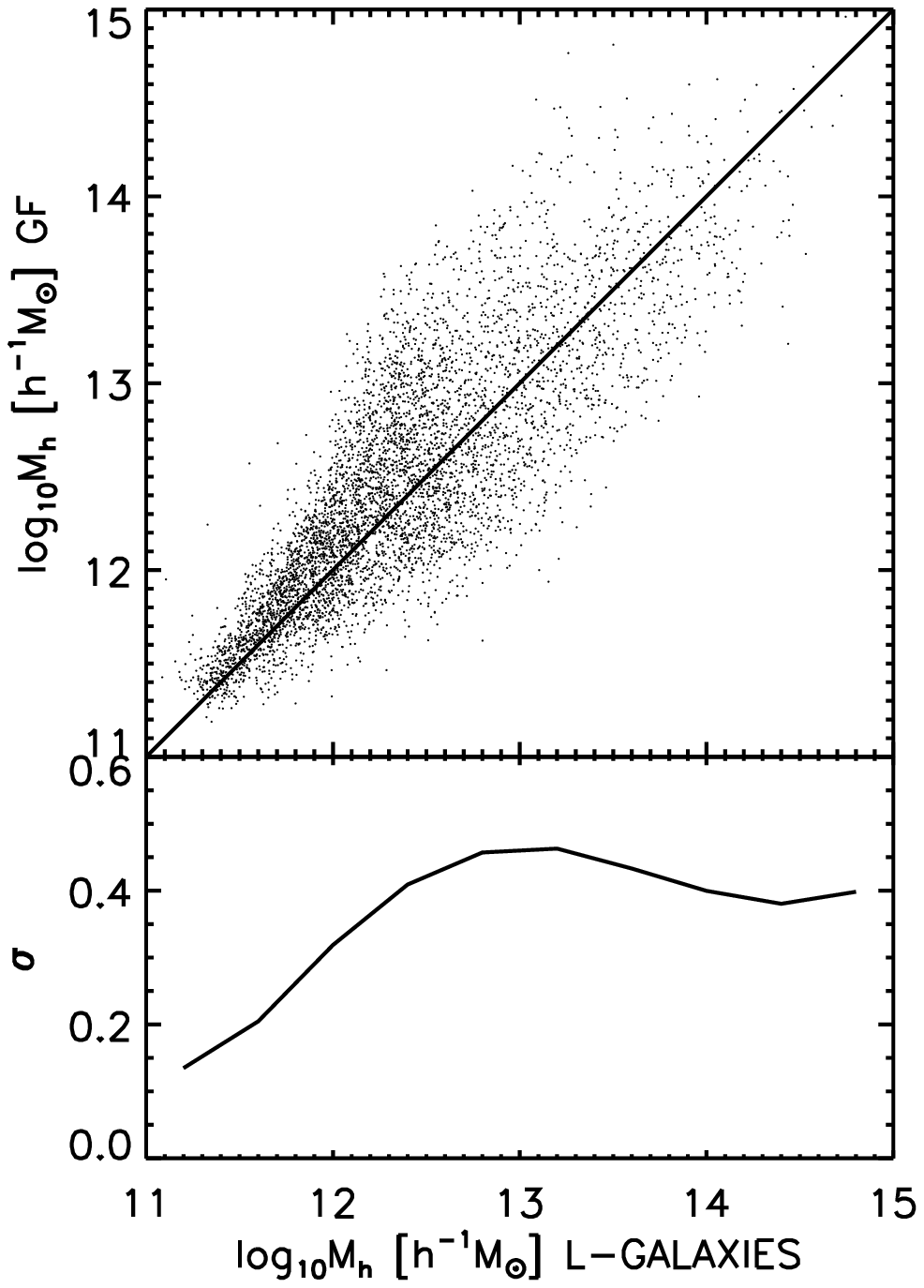,clip=true,width=0.3\textwidth}
    \epsfig{figure=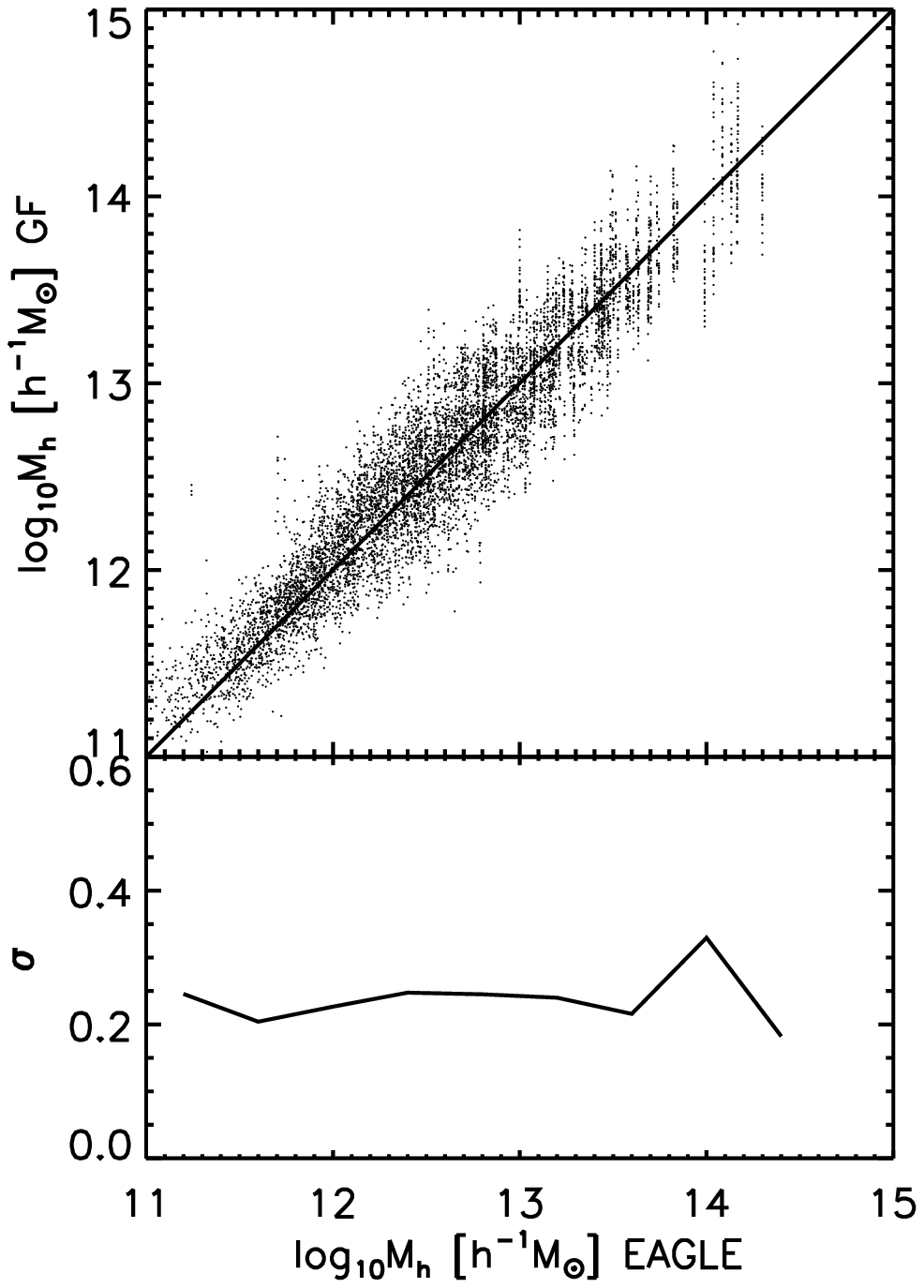,clip=true,width=0.3\textwidth}
  \end{center}
\caption{The halo mass assigned by group finder (vertical axis) versus the original halo mass
(horizontal axis) for L-GALAXIES (top left panel) and EAGLE (top right panel).
Only 20,000 galaxies randomly selected from the mock samples are shown here.
The bottom two panels show the scatter of the assigned halo mass versus
the original halo mass.
}
 \label{fig:halo_mass}
\end{figure*}


In order to better understand how and why the group finder impacts the
results, and why the impact is so much larger for L-GALAXIES than for
EAGLE, we now take a closer look at how the group finder performs for
both models.  The most relevant pieces of information provided by the
group finder are the halo masses and the central/satellite
classification. In order to better understand how the group finder
impacts our results, we therefore examine the mis-classification rate
and the reliability of the assigned halo masses in L-GALAXIES+GF and
EAGLE+GF. The upper panels of Figure \ref{fig:halo_mass} plot the halo
masses assigned by the group finder versus true halo mass, for both
L-GALAXIES and EAGLE.  The scatter in the assigned halo mass as a
function of the true halo mass is shown in the bottom panels. Overall,
there is a good linear relation between the assigned and true masses
for both L-GALAXIES and EAGLE. However, the relation given by
L-GALAXIES has significantly larger scatter, especially for massive
halos ($\sim$0.4 dex). For EAGLE, the scatter is much smaller,
$\sim$0.2 dex, and shows almost no dependence on halo mass.  We have
also checked the mis-classification rate of centrals versus
satellites, defined as the fraction of centrals/satellites in
L-GALAXIES (or EAGLE) but classified as satellites/centrals in
L-GALAXIES+GF (or EAGLE+GF).  The mis-classification rates of centrals
and satellites in L-GALAXIES+GF are 0.06 and 0.33, while they are
about 0.03 and 0.16 in EAGLE+GF.

We can thus conclude, that overall the group finder performs
significantly better in EAGLE than in L-GALAXIES. Further tests show
that the main reason for this difference is that EAGLE predicts a much
tighter relation between halo mass and the total stellar mass of its
member galaxies (see Figure \ref{fig:tot_mstar_mhalo} in the Appendix).
Since the group finder assigns halo masses under the ansatz of a
one-to-one, monotonic relation between total stellar mass and halo
mass, the poor performance of the group finder in the case of
L-GALAXIES is easily understood.

Another issue that plays an important role is the central/satellite
mis-classification, which causes centrals and satellites in the group
catalog to look more similar than they really are
\citep[][]{Campbell-15}.  Since the mis-classification is
significantly larger in L-GALAXIES than in EAGLE, the difference
between L-GALAXIES+GF and L-GALAXIES is larger than between EAGLE+GF
and EAGLE.  The fact that centrals and satellites in EAGLE are
intrinsically more similar to each other than in L-GALAXIES
(cf. Figure \ref{fig:SAM_quench_mh_1th}), adds to this enhanced effect
of the group finder in the case of L-GALAXIES.


\begin{figure*}
  \begin{center}
    \epsfig{figure=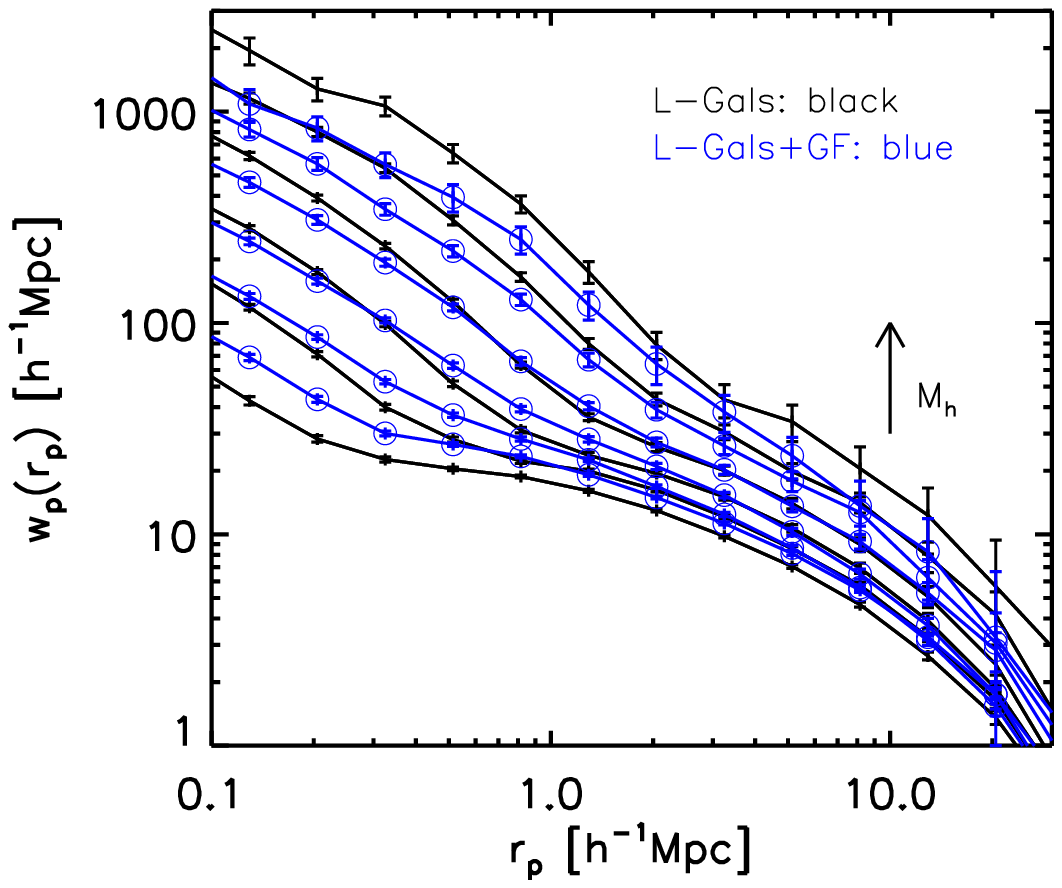,clip=true,width=0.35\textwidth}
    \epsfig{figure=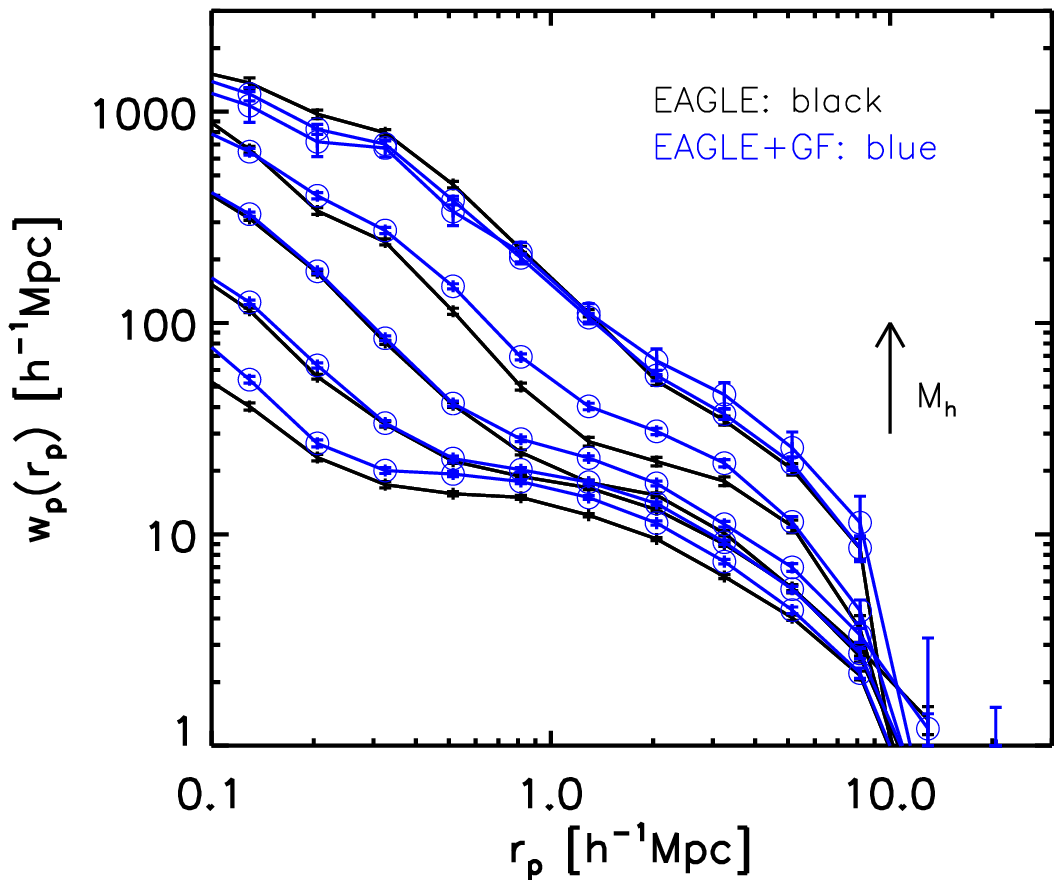,clip=true,width=0.35\textwidth}
    \epsfig{figure=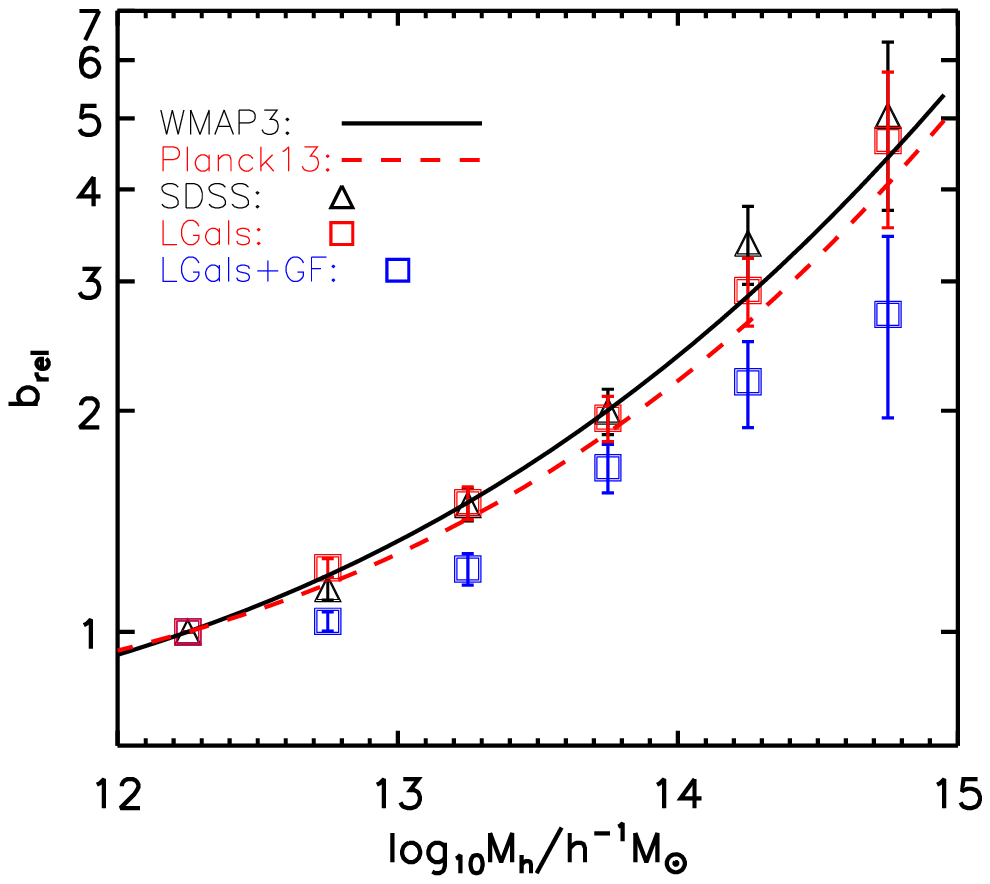,clip=true,width=0.35\textwidth}
    \epsfig{figure=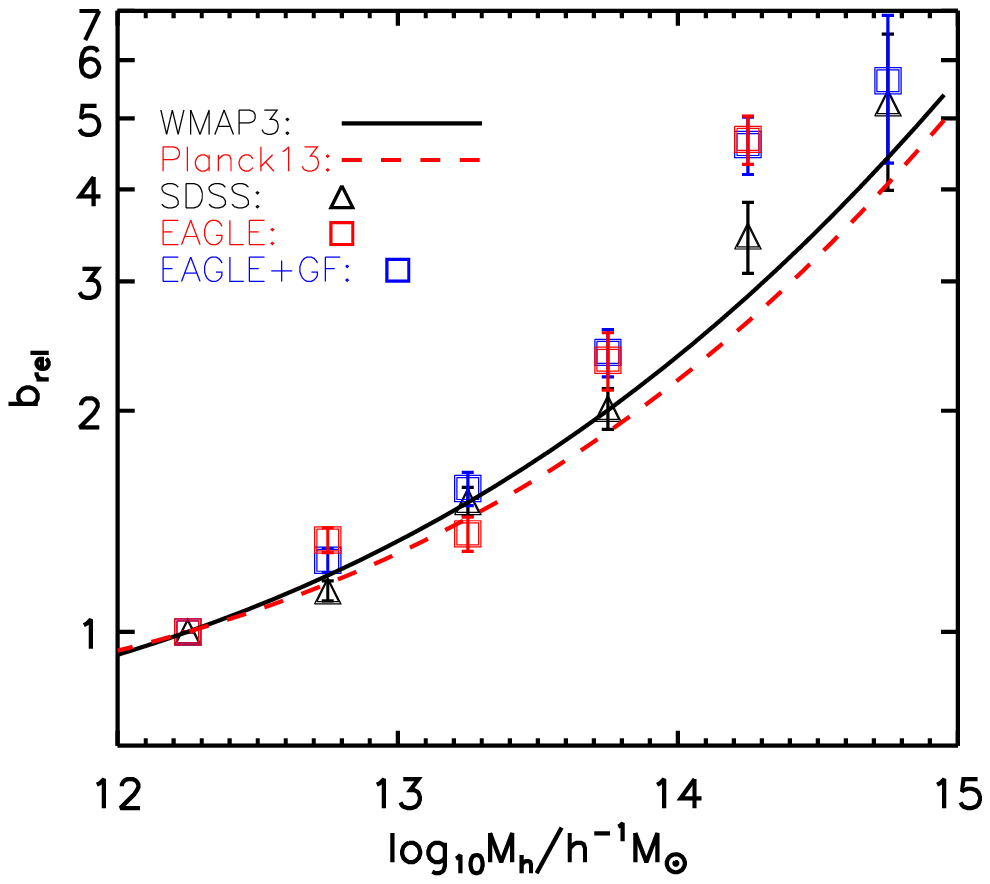,clip=true,width=0.35\textwidth}
  \end{center}
\caption{Top panels show the 2PCCF between halos and mock galaxies for
  halos of different mass. The left panel shows the results for
  L-GALAXIES (black lines) and L-GALAXIES+GF (blue lines), while the
  right panel shows the results for EAGLE (black lines) and EAGLE+GF
  (blue lines). Halos are divided into six halo mass bins, $\logmhalo
  \in[12, 12.5)$, $[12.5, 13.0)$, $[13.0, 13.5)$, $[13.5, 14.0)$,
          $[14.0, 14.5)$, and $[14.5, 15.0]$, respectively. Bottom
            panels show the relative halo bias (see text for the
            definition) as a function of halo mass. The left panel
            shows the results for L-GALAXIES (red squares) and
            L-GALAXIES+GF (blue squares), while the right panel shows
            the results for EAGLE (red squares) and EAGLE+GF (blue
            squares). The results for SDSS galaxies are shown in black
            triangles. For comparison, we also show the theoretical
            predictions of two cosmological models, Planck cosmology
            adopted by L-GALAXIES and EAGLE (red dashed line), and
            WMAP3 cosmology adopted by SDSS group catalog (black solid
            line). Errors are estimated by using bootstrap method with
            100 samples.  }
 \label{fig:halo_bias}
\end{figure*}

The difference in the performance of the group finder for L-GALAXIES
and EAGLE is a concern, as it is not known {\it a priori} whether
L-GALAXIES or EAGLE is a better representation of the real
Universe. Here we use an independent test to show that the EAGLE
simulation is more reminiscent of the real Universe than the
population of galaxies in the L-GALAXIES semi-analytical model.

As shown above, the problem of the group finder in its application to
L-GALAXIES is the large uncertainties in the assigned halo mass.  It
is well known that halo clustering depends strongly on halo mass
\citep[e.g.][]{Mo-White-96, Sheth-Mo-Tormen-01}.  Hence, if
  there are large errors in the assigned halo masses, this should
  reveal itself in the clustering properties of the galaxy groups.  In
  particular, at the massive end, where the halo bias depends strongly
  on halo mass, large errors will result in a significant reduction of
  the mass dependence of the clustering of the groups. This idea was
  tested in \cite{Wang-08}, who measured the relative bias of groups
  selected from the \cite{Yang-07} SDSS galaxy group catalog. They found that
  clustering-dependence on the inferred group mass to be in excellent
  agreement with that expected for halos in the $\Lambda$CDM
  concordance cosmology. This indicates that the errors in the
  inferred group masses have to be relatively small.  We now repeat
  this analysis of \cite{Wang-08} by measuring the relative halo bias
  as inferred from L-GALAXIES+GF, EAGLE+GF, and SDSS, and comparing
  the results with theoretical expectations.


To estimate the halo bias, for each group catalog, we divide halos
(groups) into six halo mass bins uniformly spaced in $\log(M_{\rm
  h}/h^{-1}{\rm M}_\odot)$ between $12$ and to $15$. We calculate the
projected two-point cross correlation function (2PCCF) between the
groups in each mass bin and all galaxies in the corresponding catalog.
The 2PCCFs so obtained are shown in Figure \ref{fig:halo_bias} as the
black and blues lines for L-GALAXIES/EAGLE and L-GALAXIES+GF/EAGLE+GF,
respectively. Errors are estimated using the bootstrap
method \citep[e.g.][]{Barrow-Bhavsar-Sonoda-84}, and the details for
calculating the 2PCCF can be found in \cite{Li-06}.  Note that the
results for L-GALAXIES/EAGLE are independent of the group finder,
while those for L-GALAXIES+GF/EAGLE+GF are affected by the group
finder.

The clustering amplitude increases with increasing halo mass, in
agreement with the fact that more massive halos are more strongly
clustered. For L-GALAXIES, the group finder leads to a higher 2PCCF
for the lowest halo mass bin, and a significantly lower 2PCCF for the
three highest halo mass bins ($M_{\rm h}>10^{13.5}h^{-1}$\msolar),
especially at scales $<1 h^{-1}$Mpc. For EAGLE, the difference between
EAGLE and EAGLE+GF are smaller than that between L-GALAXIES and
L-GALAXIES+GF, although significant differences are present in the
lowest halo mass bin.  The 2PCCFs of EAGLE and EAGLE+GF drops sharply
at $r_{\rm p}>10$ $h^{-1}$Mpc, which is a consequence of the small
simulation volume.

To quantify the change of the clustering amplitude with halo mass, we
estimate the relative bias, defined as the ratio between the
clustering amplitude for halos of given $M_{\rm h}$ and that for halos
of $M_{\rm h}\sim 10^{12.25}h^{-1}M_{\odot}$.  The relative bias is
calculated using the 2PCCFs in the range 4 $h^{-1}$Mpc $<r_p<$ 20
$h^{-1}$Mpc for L-GALAXIES, while that for EAGLE is calculated using
the measurements within 4 $h^{-1}$Mpc $<r_p<$ 10 $h^{-1}$Mpc to reduce
the impact of the box size. The results are shown in the bottom panels
of Figure \ref{fig:halo_bias}. For comparison, we also show the
theoretical predictions of the \cite{Sheth-Mo-Tormen-01} model for the
two cosmological models, Planck adopted by L-GALAXIES and EAGLE, and
WMAP3 adopted by \cite{Yang-07} when constructing the SDSS group
catalog.

As can be seen, the relative halo bias for L-GALAXIES and EAGLE agree
well with the theoretical curve. The halo bias obtained from
L-GALAXIES+GF is, however, much lower than that from L-GALAXIES. This
again indicates the failure of the group finder in the application to
L-GALAXIES.  In contrast, the halo bias obtained from EAGLE+GF
resembles that from EAGLE and follows well the theoretical prediction,
indicating that the group finder works successfully for EAGLE.  More
importantly, the halo bias of SDSS follows closely the theoretical
prediction, as in EAGLE, and very different from that in
L-GALAXIES. Unless our theoretical prediction for the mass dependence
of the halo bias is significantly in error, this indicates that the
scatter between halo mass and total stellar mass in the SDSS is not
significantly larger than in the EAGLE simulation, and smaller than
what is predicted by L-GALAXIES.  It also suggests that the mass
estimates for the SDSS groups are reliable, at least in a statistical
sense.

\section{Summary and discussion}
\label{sec:summary}

In Paper I, we found that the quenched fraction correlates with a
variety of parameters, such as halo mass, stellar mass, central
velocity dispersion, bulge-to-total ratio, and halo-centric distance,
and that the correlations are almost identical for central and
satellite galaxies as long as the samples are properly controlled. In
the present paper, we investigate the quenching properties of galaxies
in two galaxy formation models, L-GALAXIES and EAGLE, and examine
whether these two different models can reproduce the similarity
between the quenching properties of centrals and satellites. We mimic
the observations by constructing flux-limited mock galaxy samples from
the two galaxy formation models.  Since our main focus is on the
difference between the two populations of galaxies, rather than on the
overall trends predicted by the models, we define `quenched' galaxies
in the model samples such that they match the observed \fq-$M_*$
relation of SDSS galaxies.  The results from the two models are
presented both with and without adopting the group finder, so that the
uncertainties induced by the group finder can be examined. Our main
results are as follows:

\begin{itemize}

\item At given stellar mass, the overall \fq-$M_{\rm h}$ relations for
  L-GALAXIES agree with the observational results better than those
  for EAGLE.  However, the L-GALAXIES model predicts significantly
  higher quenched fractions for centrals than for satellites at fixed
  halo mass and stellar mass. In contrast, the differences between
  centrals and satellites in the EAGLE simulation are much smaller.

\item The L-GALAXIES model predicts very different \fq-$M_{\rm BH}$
  relations for centrals and satellites at given halo mass and stellar
  mass. In particular, the predicted quenched fraction for centrals
  changes rapidly from about zero to about one around $M_{\rm
    BH}\sim10^{6.5}\Msun$, while the quenched fraction of satellites
  reveals a much weaker dependence on $M_{\rm BH}$. In contrast, the
  EAGLE simulation predicts \fq-$M_{\rm BH}$ relations for centrals
  and satellites that are very similar, when controlling for both
  stellar and halo mass.

\item The L-GALAXIES model fails to reproduce the observed
  independence of \fq\ on halo-centric distance for massive galaxies
  at low to intermediate halo mass ($12.0 < \log(M_{\rm h}/\Msunh) <
  13.2$). EAGLE, on the other hand, matches the observations at these
  halo mass bins better.

\item After applying group finder, the results for EAGLE only change
  slightly.  In contrast, the group finder significantly reduces the
  differences between centrals and satellites in L-GALAXIES,
  especially for the dependence of \fq\ on halo mass. However,
  significant differences between the two populations remain, in
  particular regarding the dependence of \fq\ on black hole
  mass.

\item Overall, the group finder works better on EAGLE than on
  L-GALAXIES, yielding better halo mass assignments and better
  central-satellite identification. This indicates that the
  performance of the group finder depends significantly on the
  detailed predictions of galaxy formation models. As an additional
  test of the performance of the group finder, we examined the
  clustering of the groups identified by the group finder. In the case
  of L-GALAXIES, the inferred clustering as a function of group mass is
  inconsistent with the expected mass dependence of the halo bias.
  EAGLE, on the other hand, is similar to the SDSS data, in that the
  inferred clustering of the groups is in excellent agreement with
  expectations based on the halo bias. This suggests that the amount
  of scatter in the relation between halo mass and total stellar mass
  in EAGLE is in agreement with that in the SDSS data, while
  L-GALAXIES predicts a scatter that is too large. In addition, the
  fact that the measured halo-bias for the SDSS groups agrees well
  with the theoretical expectations suggests that the results inferred
  from the SDSS group catalog are reliable.

\end{itemize}

As we have shown, centrals and satellites in L-GALAXIES have very
different quenching properties, even when stellar mass and halo mass
are fixed. This difference is likely caused by the fact that
L-GALAXIES, and SAMs in general, treat centrals and satellites
differently. In particular, SAMs typically include a number of
processes that operate only on satellite galaxies. In early versions
of various SAMs, it was assumed that satellite-specific processes
strip satellite galaxies of their hot gas reservoir as soon as they
become a satellite galaxy, i.e., as soon as they are accreted by a
halo more massive than their own
\citep[e.g.][]{DeLucia-Kauffmann-White-04, Kang-05, Croton-06}.  As
first pointed out in \citet{Weinmann-06}, this leads to a rapid
decline of star formation and a fraction of red or quenched satellites
that is much higher than observed. Subsequent SAMs therefore adopted a
more gentle stripping of the hot gas \citep[e.g.][]{Font-08,
  Kang-vandenBosch-08, Weinmann-10, Guo-11}.  In the more recent
L-GALAXIES SAM of \citet{Henriques-15}, ram-pressure is assumed to
only operate in the most massive halos. Although these modifications
reduce the difference between centrals and satellites, the results
presented here suggest that the remaining differences may still be too
large.


Another origin for the large differences between centrals and
satellites in L-GALAXIES may come from the particular treatment of AGN
feedback, which is assumed to be proportional to both black hole mass
and the mass of the surrounding hot gas. For centrals, this results in
a quenched fraction as function of black hole mass that is almost
step-function-like, transiting from close to zero to close to unity
around $M_{\rm BH} \sim 10^{6.5}M_{\odot}$ (see
Figure \ref{fig:SAM_quench_bh_2th}). The quenched fraction of satellite
galaxies, on the other hand, reveals a much weaker dependence on black
hole mass. We have examined the hot gas mass fractions of centrals and
satellites in L-GALAXIES of similar stellar mass and halo mass. In
general, central galaxies have much more hot gas than satellites. More
importantly, the amount of hot gas in centrals varies little from
galaxy-to-galaxy (at fixed stellar mass), whereas the hot gas mass
fraction of satellites spans several orders of magnitude. As a
consequence, the impact of AGN feedback on centrals is almost entirely
regulated by the mass of the black hole. For satellites, on the other
hand, the much broader distribution of hot gas mass fractions explains
the much weaker dependence on black hole mass.

Contrary to L-GALAXIES, the EAGLE simulation nicely reproduces the
similarities between centrals and satellites regarding correlations of
the quenched fraction with various other quantities. The simulation
may, therefore, provide useful insights into the various processes
that cause galaxies to quench their star formation.  Whereas
L-GALAXIES assume that the AGN feedback feedback efficiency depends on
the total mass of hot gas, in EAGLE the amount of feedback energy is
assumed to depend on the {\it local} density and temperature of the
gas surrounding the central black hole. Since these local gas
properties are likely to be less strongly affected by the environment
than the total hot gas mass, this may explain why centrals and
satellites in EAGLE appear more similar.

The comparison with SDSS data presented here suggests that the
quenching properties of galaxies in EAGLE are in better agreement with
the data than those in L-GALAXIES. However, we caution that
the results based on the group finder can cause centrals and satellite
to appear more similar than they are in reality. Indeed, after applying
the halo-based group finder of \citet{Yang-05} over mock data extracted
from L-GALAXIES, some of the dramatic differences between centrals and
satellites present in the SAM are drastically suppressed. Nevertheless,
significant differences with respect to the SDSS data remain, and an
analysis of the clustering properties of the L-GALAXIES groups clearly
suggests that it is far more significantly impacted by group finder errors
than either EAGLE or SDSS. We are therefore cautiously confident that
the EAGLE simulations better captures the physics of galaxy quenching
than the L-GALAXIES semi-analytical model.

\acknowledgments

This work is supported by the National Basic Research Program of China
(973 Program)(2015CB857002, 2018YFA0404503, 2015CB857004), the
National Natural Science Foundation of China (NSFC, Nos.
11522324,11733004, 11421303, 11433005, 11621303 and 11320101002), and the
Fundamental Research Funds for the Central Universities.  EW
acknowledges the support from the Youth Innovation Fund by University
of Science and Technology of China (No. WK2030220019) and China
Postdoctoral Science Foundation funded project (No. BH2030000040). SHC
is supported by the Fund for Fostering Talents in Basic Science of the
National Natural Science Foundation of China NO.J1310021. HM also
acknowledges the support from NSF AST-1517528 and NSFC-11673015.  FvdB
is supported by the US National Science Foundation through grant AST
1516962, and by the National Aeronautics and Space Administration
through Grant No. 17-ATP17-0028 issued as part of the Astrophysics
Theory Program.

\bibliography{rewritebib.bib}

\appendix
\counterwithin{figure}{section}
\section{A: Defining the population of quenched galaxies} \label{sec:appendix}

\begin{figure*}
  \begin{center}
    \epsfig{figure=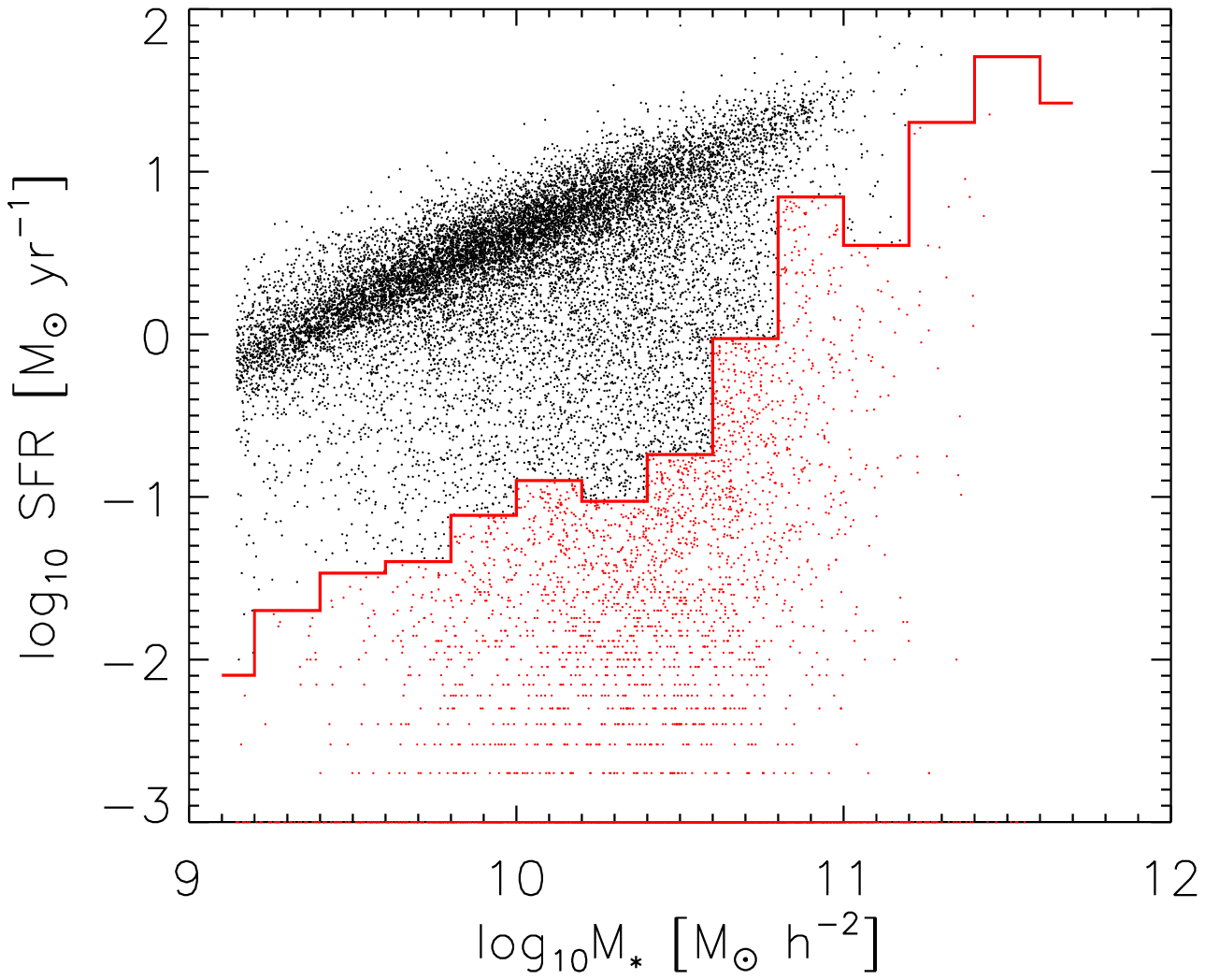,clip=true,width=0.45\textwidth}
    \epsfig{figure=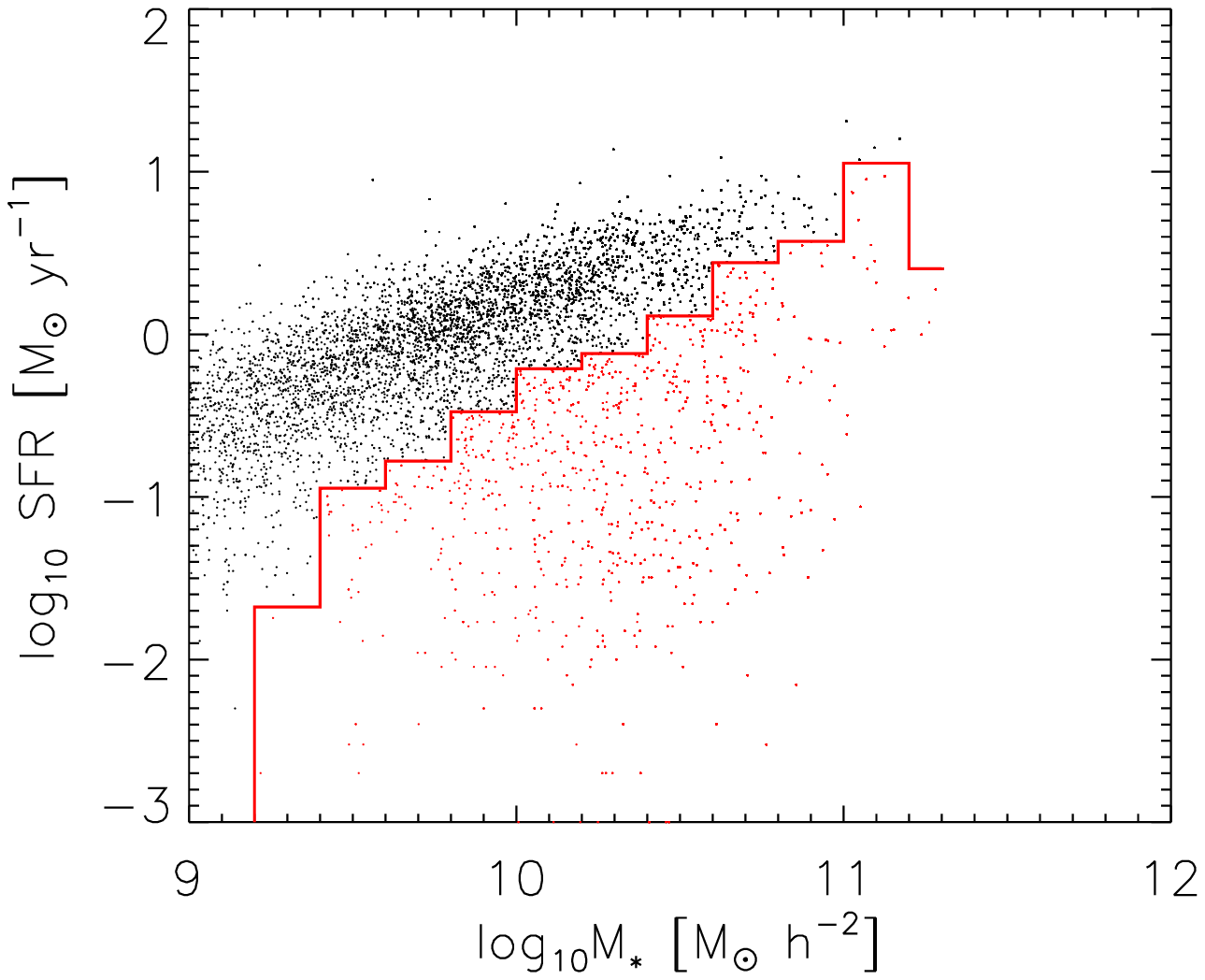,clip=true,width=0.45\textwidth}
  \end{center}
\caption{The SFR-$M_*$ relation for L-GALAXIES (left panel) and EAGLE (right panel). The red demarcation line indicates the SFR threshold, which is used to separate star-forming (black dots) from quenched galaxies (red dots). In displaying the SFR-$M_*$ relation, we randomly select 20,000 galaxies from each sample. Note that many quenched galaxies with zero SFR are not shown here, as they fall below the scale of the plots.}
\label{fig:def_quench_threshold}
\end{figure*}

\begin{figure*}
  \begin{center}
    \epsfig{figure=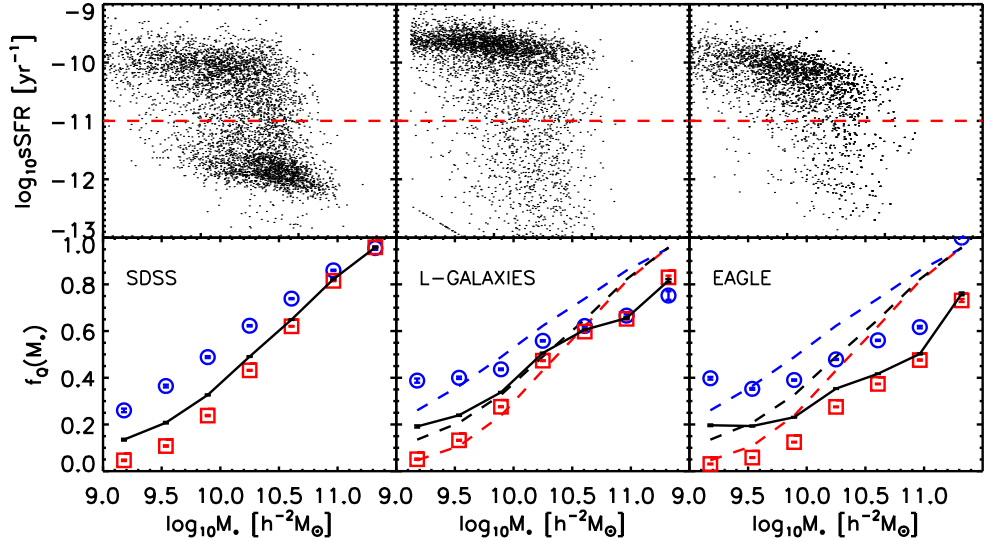,clip=true,width=0.8\textwidth}
  \end{center}
\caption{The sSFR-$M_*$ relation and \fq-$M_*$ relation for galaxies
  from SDSS (left panels), L-GALAXIES (middle panels) and EAGLE (right
  panels). In displaying the sSFR-$M_*$ relation, we randomly select
  20,000 galaxies from each sample. The red dashed line
  (sSFR=$10^{-11}yr^{-1}$) in the top panels are used to separate SF
  and quenched populations.  In each bottom panel, the blue circles,
  red squares and black line represent the \fq-$M_*$ relations for
  satellites, centrals and all galaxies, respectively.  For
  comparison, we display the \fq-$M_*$ relations for SDSS galaxies in
  dashed lines in the bottom-middle and bottom-right panels. }
 \label{fig:def_quench_mstar_0th}
\end{figure*}

To allow for a meaningful comparison of the quenching fractions in the
data and the models, we define the quenched fractions in L-GALAXIES
and EAGLE such that they match the observed \fq-$M_*$ relation of SDSS
galaxies. Figure \ref{fig:def_quench_threshold} shows the SFR-$M_*$ relation
for the SF and quenched galaxies in L-GALAXIES and EAGLE.  The red demarcation
line indicates the SFR thresholds, below which galaxies are considered quenched.
This definition is not commonly used in the literature, but is adopted here so that a 
meaningful comparison between models and observational data can be made. 
Note that  a galaxy defined as quenched in this way, may
still be actively forming stars in the L-GALAXIES or EAGLE
simulations.

To get an overall impression of the \fq-$M_*$
relation predicted by the models, here we present results based on the 
commonly used definition of quenched population for all the three samples, 
L-GALAXIES, EAGLE, and SDSS. 
The upper panels of Figure \ref{fig:def_quench_mstar_0th}
plot the sSFR versus stellar mass for galaxies in the SDSS (left-hand
panel), the L-GALAXIES model (middle panel) and the EAGLE simulations
(right-hand panel). Whereas the SDSS data reveals a very pronounced
population of quenched galaxies, this is not the case for L-GALAXIES
or EAGLE. This is mainly because the models predict many galaxies with
zero SFR, which falls below the scale of the plots.  If we split the
populations in quenched and star-forming at a sSFR of
$10^{-11}yr^{-1}$, indicated by the dashed lines in the top panels, we
obtain the quenched fractions as function of stellar mass indicated in
the lower panels. Both L-GALAXIES and EAGLE reproduce the global
trends seen in the SDSS data of \fq\ increasing with mass, and of
satellites (blue circles) having higher quenched fractions than
centrals (red squares) at fixed stellar mass.  However, L-GALAXIES
under-predicts the quenched fraction of galaxies at the massive end
(\lgmstar$>$10.6). The EAGLE simulation fairs even worse in
reproducing the detailed \fq-$M_*$ relations observed in the SDSS. In
order for our results not to be affected by these overall differences,
at each stellar mass bin we define the sSFR below which galaxies are
defined to be quenched such that both L-GALAXIES and EAGLE yield
\fq-$M_*$ relations relations that are identical to the SDSS results
(see \S\ref{sec:mockcat}) for details).

\section{B: Controlled stellar mass ranges for EAGLE (EAGLE+GF) and L-GALAXIES (L-GALAXIES+GF)}
\begin{figure*}
  \begin{center}
    \epsfig{figure=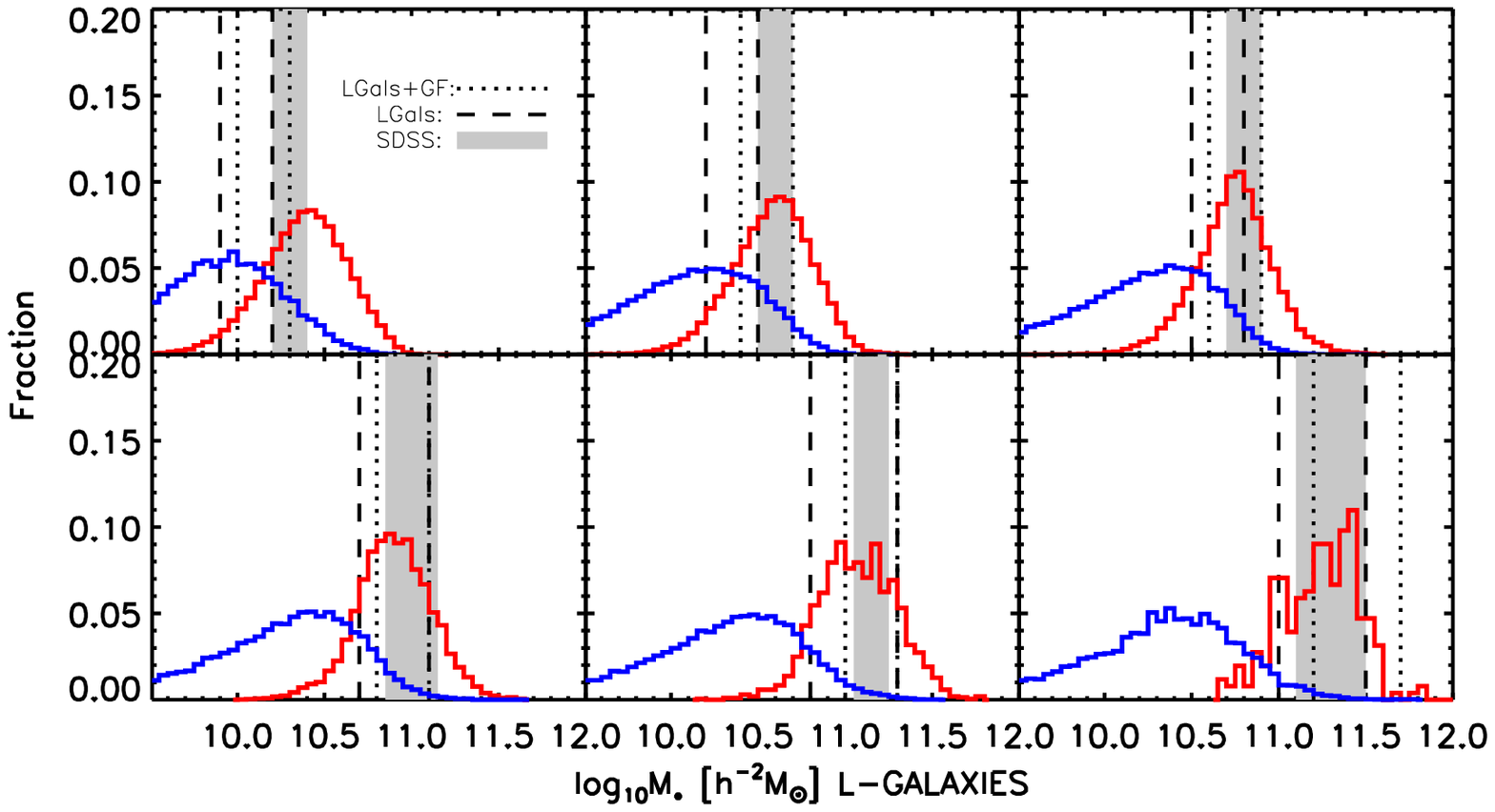,clip=true,width=0.7\textwidth}
    \epsfig{figure=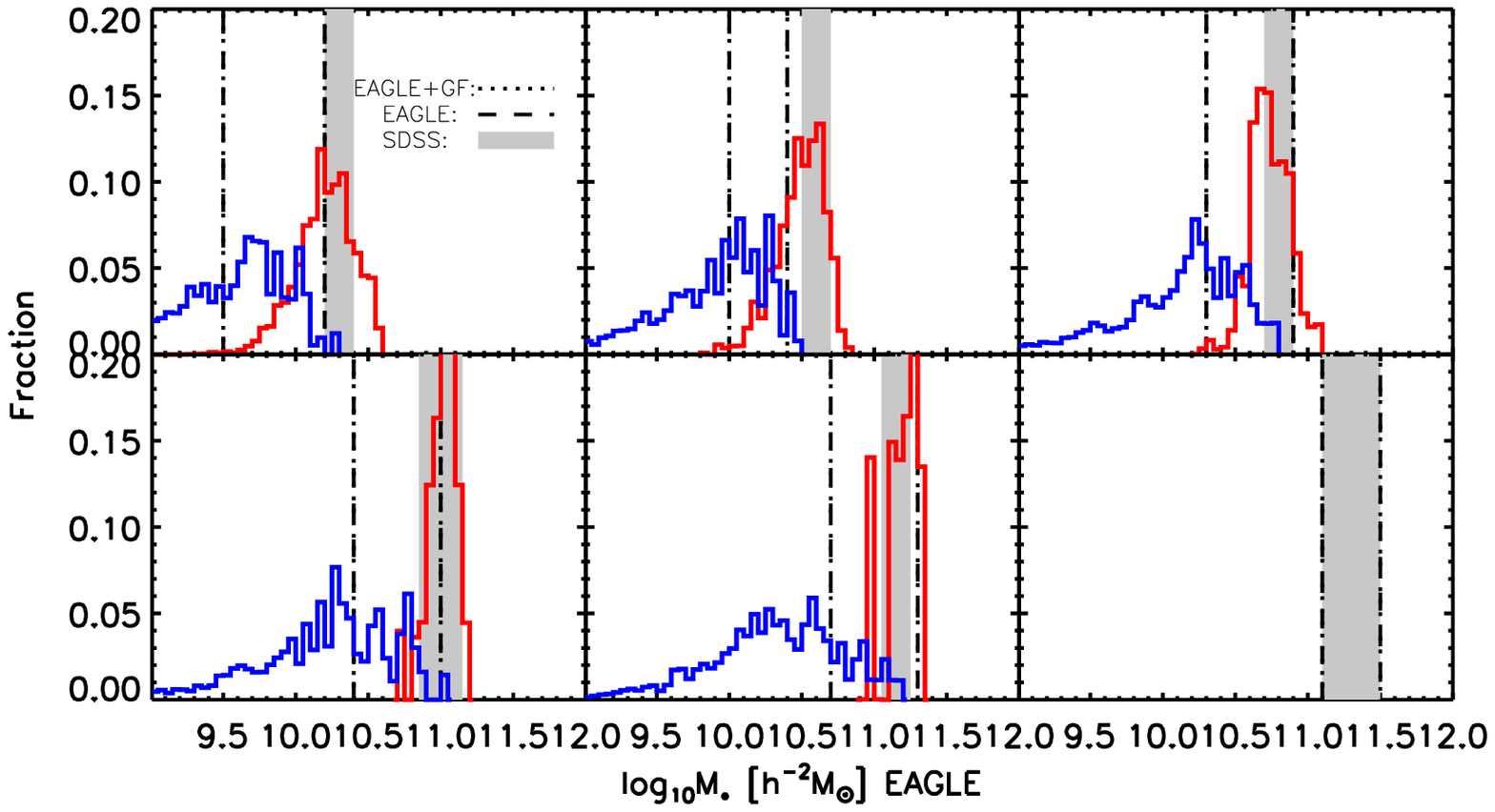,clip=true,width=0.7\textwidth}
  \end{center}
\caption{The normalized stellar mass distribution of centrals and
  satellites for the subsamples of L-GALAXIES (top panels) and EAGLE
  (bottom panels). In each panel, the red and blue histograms
  represent the stellar mass distribution of centrals and satellites,
  respectively.  The controlled stellar mass ranges for SDSS,
  L-GALAXIES (EAGLE) and L-GALAXIES+GF (EAGLE+GF) are displayed in
  shaded region, the dashed lines and the dotted lines, respectively.}
 \label{fig:mass_dis_control}
\end{figure*}

In order to compare the \fq-$M_{\rm BH}$ relations of centrals and
satellites, we have to control both stellar mass and halo
mass. However, it is known that centrals and satellites occupy
different loci in the stellar mass and halo mass plane. In Figure
\ref{fig:mass_dis_control}, we show the stellar mass distributions of
centrals and satellites in various halo mass bins for L-GALAXIES (top
group of panels) and EAGLE (bottom group of panels). The distributions
for SDSS galaxies can be found in figure 5 of paper I. The overlap
between the two populations is usually narrow and varies with halo
mass. For each halo mass bin, the stellar mass range is manually
selected to make sure that there are a large enough number of centrals
and satellites to facilitate a meaningful comparison. The stellar mass
ranges for L-GALAXIES, L-GALAXIES+GF, EAGLE, and EAGLE+GF are marked
in Figure \ref{fig:mass_dis_control}. For comparison, the shaded
regions show the corresponding stellar mass ranges for SDSS
galaxies. For L-GALAXIES, the halo masses assigned by the group finder
are quite different from the true halo masses (see Figure
\ref{fig:halo_mass}), which explains why the selected stellar mass
ranges for L-GALAXIES and L-GALAXIES+GF are somewhat different. Since
the stellar mass distributions in each halo mass bin are different
among L-GALAXIES, EAGLE and SDSS, the chosen stellar mass bins for the
two models and the observations are also different.  As one can see,
the controlled stellar mass ranges for SDSS galaxies are higher than
those for L-GALAXIES and EAGLE, especially in less massive halos
($M_{\rm h}<h^{-1}10^{13.5}$\msolar). The differences are reduced at
the massive end.  Specifically, the controlled stellar mass ranges of
L-GALAXIES+GF are more consistent with those of SDSS than
L-GALAXIES. The controlled stellar mass ranges are the same for EAGLE
and EAGLE+GF, which are systematically lower than those of SDSS. The
reader should keep these differences in mind when interpreting the
results in Figures~\ref{fig:SAM_quench_bh_2th} and~\ref{fig:SAM_quench_bh_2th_GF}.

\section{C: The total stellar mass of member galaxies versus the halo mass of groups for L-GALAXIES and EAGLE}

\begin{figure*}
  \begin{center}
    \epsfig{figure=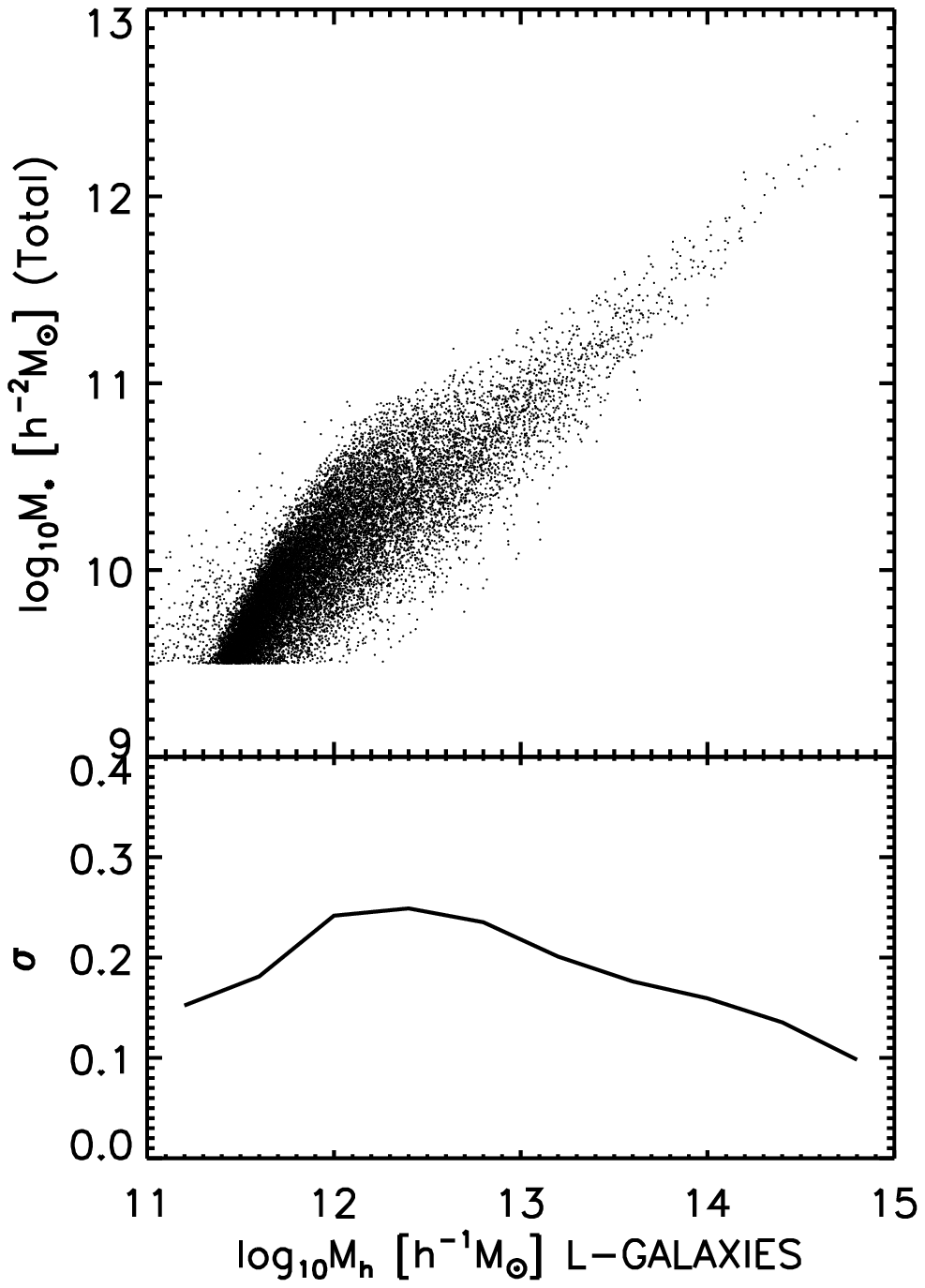,clip=true,width=0.3\textwidth}
    \epsfig{figure=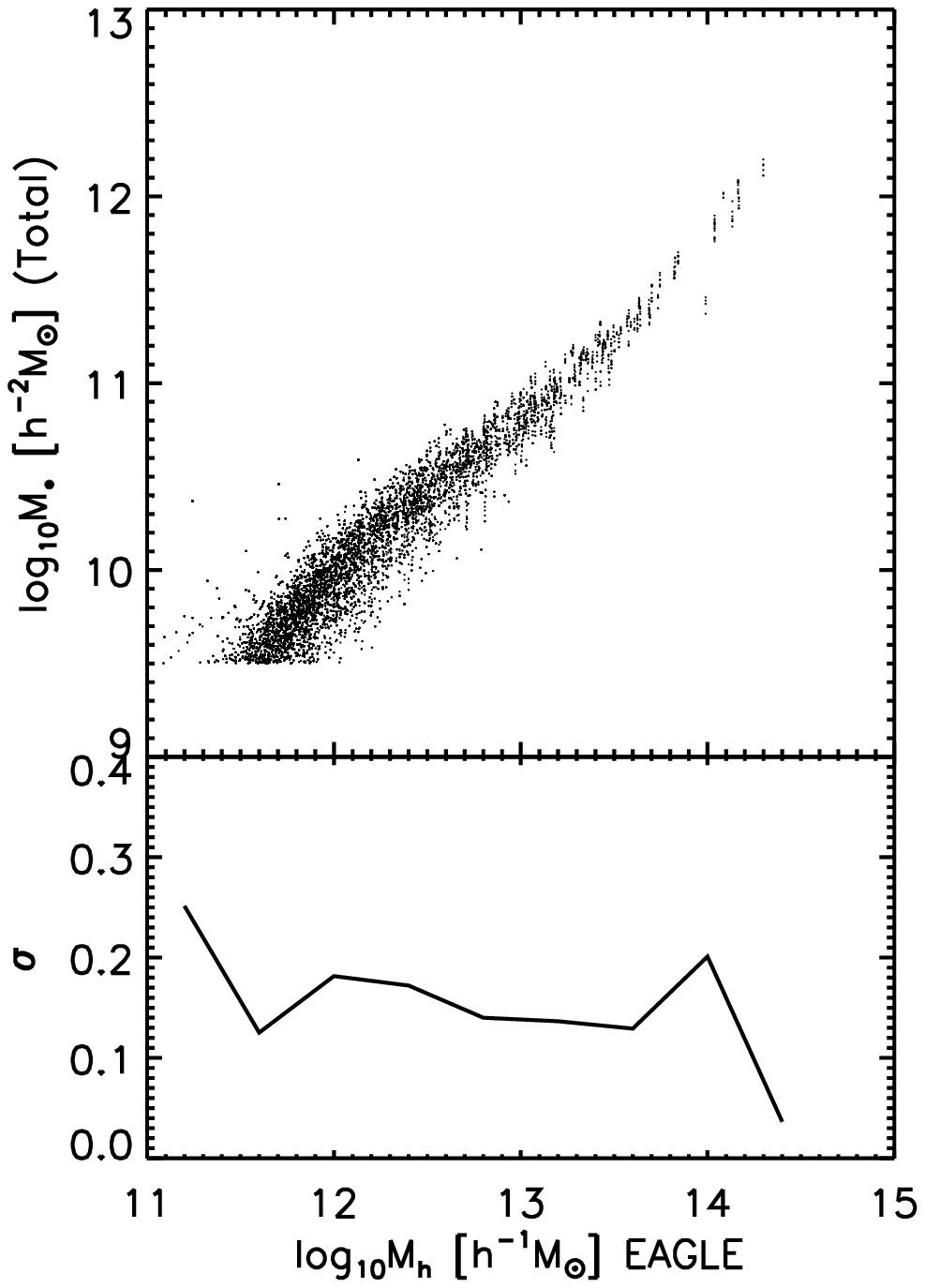,clip=true,width=0.3\textwidth}
  \end{center}
\caption{ The total stellar mass of member galaxies versus the halo
  mass of groups for L-GALAXIES (top left panel) and EAGLE (top right
  panel).  Only 20,000 groups are randomly selected to be shown here.
  The bottom two panels show the scatter of total stellar mass as a
  function of the halo mass. Note that the halo mass and total stellar
  mass of member galaxies presented here are directly taken or derived
  from L-GALAXIES and EAGLE, without applying group finder algorithm.
}
 \label{fig:tot_mstar_mhalo}
\end{figure*}

As shown in Figure~\ref{fig:halo_mass}, for the L-GALAXIES model
  there is a large scatter (up to $\sim$0.4 dex) between the real halo
  mass and the halo mass assigned by the group finder. This scatter is
  significantly smaller ($\sim$0.2 dex) in the case of the EAGLE
  simulations. The group finder assign halo masses to each group
  assuming a tight correlation between halo mass and the total stellar
  mass of its member galaxies. Hence, the fact that the masses
  assigned to the groups in EAGLE are more accurate than in the case
  of L-GALAXIES is likely to have its origin in a larger scatter
  between halo mass and total stellar mass in L-GALAXIES. To test
  this, Figure \ref{fig:tot_mstar_mhalo} plots the relationship between
  the total stellar mass, $M_{\rm *,tot}$, of member galaxies and halo
  mass for groups in both L-GALAXIES (left) and EAGLE (right). The
  lower panels plot the corresponding scatter in $\log(M_{\rm *,
    tot})$ as function of halo mass. Overall, L-GALAXIES indeed
  reveals a larger amount of scatter, reaching $\sim 0.25$ dex at
  $M_{\rm h}\sim10^{12.4}h^{-1}$\msolar. In contrast, the scatter in
  EAGLE is only $\sim 0.16$ dex around that mass scale. Hence, we
  conclude that the main origin for the larger errors in the assigned
  group masses in L-GALAXIES is a larger scatter in the relation
  between halo mass and total stellar mass of its member galaxies.

\clearpage
\label{lastpage}
\end{document}